# Nanowire photodetectors based on wurtzite semiconductor heterostructures

**Maria Spies[1] and Eva Monroy[2]**

[1] University Grenoble-Alpes, CNRS, Institut Néel, 25 av. des Martyrs, 38000 Grenoble, France

[2] University Grenoble-Alpes, CEA, INAC-PHELIQS, 17 av. des Martyrs, 38000 Grenoble, France

OrcIDs:

Maria Spies: 0000-0002-3570-3422

Eva Monroy: 0000-0001-5481-3267

## Abstract

Using nanowires for photodetection constitutes an opportunity to enhance the absorption efficiency while reducing the electrical cross-section of the device. They present interesting features like compatibility with silicon substrates, which offers the possibility of integrating detector and readout circuitry, and facilitates their transfer to flexible substrates. Within a nanowire, it is possible to implement axial and radial (core-shell) heterostructures. These two types can be combined to obtain three-dimensional carrier confinement (dot-in-a-wire). The incorporation of heterostructures in nanowire photodetectors opens interesting opportunities of application and performance improvement. Heterojunctions or type-II heterostructures favor the separation of the photogenerated electrons and holes, and the implementation of quantum dots in a nanowire can be applied to the development of quantum photodetectors. This paper provides a general review of latest progresses in nanowire photodetectors, including single nanowires and heterostructured nanowires.

Keywords: nanowire, photodetector, wurtzite, heterostructure

Contents

# 1. Introduction

Semiconductor nanowires are quasi-one-dimensional nanocrystals, with two dimensions in the 10-100 nm range, and lengths that are several times longer, up to several micrometers. Such nano-objects constitute the last miniaturization limit for a number of electronic, optoelectronic and sensor devices [1–3]. Both single nanowires and nanowire arrays are studied as potential photodetectors using a variety of materials [4,5]. The low dimensionality, crystalline quality and high surface-to-volume ratio are important elements to understand their performance.

Nanowire photodetectors constitute an interesting approach to get around the efficiency–speed trade-off. The low electrical cross-section of nanowires implies low electrical capacitance, but this comes without degradation of total light absorption due to antenna effects. Indeed, nanowire arrays can exhibit higher absorption than a thin film of the equivalent thickness [6–8]. Another interesting feature of nanowire photodetectors is their compatibility with silicon technology, either as growth support or carrier wafer with easy transfer procedures, which opens interesting possibilities of integrating the detector and readout. Along these lines, it is particularly attractive the possibility of growth or transfer into flexible materials [9,10], which opens perspectives for the development of wearable devices.

Within a nanowire, it is possible to implement axial and radial (core-shell) heterostructures. Axial heterostructures present heterointerfaces perpendicular to the nanowire growth axis, which is the largest of the three dimensions of the nano-object. In such a configuration, lattice mismatched materials can be epitaxially grown without defects thanks to the relaxation of misfit strain by surface deformation [11], which leads to a significantly larger critical thickness in comparison to heterostructures in two-dimensional layers. This widens the possible material combinations and thicknesses in the active region of the detector. On the other hand, radial heterostructures are interesting, for instance, to provide a passivating envelope that reduces the sensitivity of the nanowire to the chemistry of the environment. Axial and radial heterostructures can be combined to obtain three-dimensional carrier confinement, which is interesting for controlling the carrier relaxation time [12,13] or reducing the effect of surface recombination [14,15].

The incorporation of heterostructures in nanowire photodetectors opens up interesting opportunities of application and performance improvement. Heterojunctions or type-II heterostructures serve the separation of the photogenerated electrons and holes, which is a requirement for the implementation of solar cells and photodetector designs that require zero-bias operation. Moreover, nanometer-size heterostructures give rise to quantum-confined energy levels, which are the base of quantum photodetectors, such as quantum-well-based or quantum-dot-based designs for resonant detection in the infrared spectral region.

In this review paper, we focus on wurtzite nanowire heterostructures. Wurtzite is the most stable crystallographic configuration of III-nitride semiconductors (GaN, AlN, InN) and ZnO. Even though MgO and CdO present cubic rock-salt structure as their most stable arrangement, it is possible to synthesize wurtzite $Mg_xZn_{1-x}O$ (x < 0.37) and $Cd_xZn_{1-x}O$ (x < 0.37) on ZnO, covering a wide range of band gaps (2.3–4.0 eV) and band offsets for the fabrication of heterostructured devices [16]. Wurtzite ZnS, ZnSe and ZnTe can also be obtained, although their most stable configuration is cubic zinc-blende. Finally, typical zinc-blende III-V compounds (III-arsenides and III-phosphides) have a tendency to grow in the wurtzite phase under the growth conditions required for nanowire formation. In these materials, the proximity in energy of the zinc-blende and wurtzite structures generally results in strong polytypism.

In wurtzite crystals, the absence of a center of symmetry leads to piezo- and pyroelectricity. The effect of these properties becomes particularly relevant in heterostructures, where the difference of polarization in the materials results in fixed charges at the heterointerfaces and therewith internal electric fields. Such phenomena can be engineered to enhance the performance of the detectors, e.g. introducing an electric field that separates electrons and holes and reduces the dark current. However, it can also be deleterious for their performance, e.g. reducing the escape probability of the excited carriers in quantum wells or introducing a potential barrier to the diffusion of minority carriers in heterojunctions.

The paper starts with an introduction to photodetectors and wurtzite nanowires, which includes a presentation of photodetector figures of merit, a description of the particularities of light coupling to single nanowires or nanowire ensembles, a brief overview of the nanowire growth methods, a summary of the structural properties associated to the wurtzite crystal symmetry and a discussion on the crystallographic growth axis and facets that are common for the various wurtzite materials. Then, chapter 2 presents the characteristic features of metal-semiconductor-metal nanowire photodetectors, where the semiconductor is homogeneous, without heterostructures. They can be either symmetric resistors or devices with a potential asymmetry due to the nature of the metal-semiconductor contacts. Chapter 3 reviews reported results on heterostructure nanowire photodetectors, differentiating radial heterostructures, axial heterostructures for band-to-band detection, axial heterostructures for intersubband detection, and heterojunctions. The paper finishes with a brief summary and a discussion on the perspectives of nanowire photodetectors.

*1.1 Photodetector figures of merit*

*1.1.1 Responsivity and gain*

The main parameter allowing to quantify the performance of any photodetector is the responsivity ($R_\lambda$), which expresses the generated photocurrent ($I_{ph}$) per incident optical power ($P_{opt}$):

$$R_\lambda = \frac{I_{ph}}{P_{opt}} = \frac{I_{ph}}{\Phi_{ph} A_{opt}} \tag{1}$$

where $\Phi_{ph}$ is the incident photon flux density, and $A_{opt}$ is the illuminated area of the detector. The expression for the responsivity can be further developed to take into account losses described with the help of the quantum efficiency and the gain. The quantum efficiency ($\eta$) quantifies the number of generated electron-hole pairs per incoming photon, whereas gain ($g$) refers to the number of actually detected electrons per generated electron-hole pair. Additionally, expressing the incident photon flux density in terms of the wavelength ($\lambda$) and natural constants, we obtain:

$$R_\lambda = \frac{I_{ph}}{\Phi_{ph} A_{opt}} = \eta g \frac{e\lambda}{hc} \tag{2}$$

where $e$ is the fundamental charge, $h$ is Planck's constant, and $c$ the speed of light.

The concept of photoconductive gain deserves further analysis. The gain can be written as:

$$g = \frac{I_{ph}/e}{P_{opt}/\left(\frac{hc}{\lambda}\right)} \tag{3}$$

In a photoconductor, i.e. a biased detector that reacts to illumination by changing its conductivity, the photocurrent can be described as $I_{ph} = V_B \Delta G$, where $V_B$ is the bias voltage and $\Delta G$ is the change in the conductance. In turn, $\Delta G$ can be expressed as:

$$\Delta G = \Delta\left(en\mu \frac{A}{l}\right) = \frac{e\mu}{l}(A\Delta n + n\Delta A) \tag{4}$$

with $n$ being the number of charge carriers, $\mu$ the carrier mobility, $A$ the cross-section of the conductance channel, and $l$ its length. The term $\Delta n$ describes a change of the number of charge carriers and is therefore linear with increasing incident optical power (note that $\Delta n = \eta \Phi_{ph}$). However, the term $\Delta A$ is accounting for changes in the cross-sectional area of the conductance channel which depends nonlinearly on the density of carriers and on the location of the Fermi level at the nanowire sidewalls. Note that equation 4 assumes no changes in the carrier mobility due to illumination.

*1.1.3 Quantum efficiency, external quantum efficiency, internal quantum efficiency, conversion efficiency*

For a photovoltaic detector (device operated at zero bias), the quantum efficiency (η) is the ratio of extracted free charge carriers (electrons in the external circuit) to incident photons. We generally differentiate between external quantum efficiency (EQE), which considers all photons impinging on the device, and internal quantum efficiency (IQE), which considers only photons that are not reflected.

The quantum efficiency (η, EQE or IQE) is a function of the photon wavelength, hence it is generally expressed as a spectral function. It should be differentiated from the "conversion efficiency" or just "efficiency" that characterizes a solar cell. In this case, efficiency refers to the ratio of electrical energy extracted from the solar energy impinging onto the cell. This parameter measures not only the performance of the solar cell as a photon-to-electron converter, but also the adaptation of the spectral response to the solar spectrum. The conditions under which the solar cell efficiency is measured must be carefully controlled, e.g. terrestrial solar cells are measured under standard AM1.5 conditions (ISO 9845-1:1992) and at a cell temperature of 25°C (contrary to solar cells used in space, which are characterized using AM0 conditions).

*1.1.4 Spectral selectivity*

When the spectral selectivity is considered as a factor of merit, the use of direct band gap materials is generally preferred. In the case of ultraviolet photodetectors, a spectral figure of merit is the ultraviolet-to-visible rejection ratio, which is calculated as the peak responsivity in the ultraviolet range divided by the responsivity at 400 nm.

The spectral selectivity can be enhanced by integrating the active region into a resonant optical cavity, forming what is called resonant-cavity-enhanced photodetectors [17]. In addition to the spectral selectivity, the optical cavity serves to enhance the device responsivity without increasing the size of the active region.

*1.1.5 Detectivity*

The minimum detectable optical power, i.e. the optical power that provides a signal-to-noise ratio equal to one, is called the noise equivalent power, *NEP*. The NEP can be calculated by dividing the detector noise current ($I_N$) by the responsivity. The detectivity, *D*, is defined as the inverse of the *NEP*:

$$D = \frac{1}{NEP} = \frac{R_\lambda}{I_N} \quad (5)$$

The parameter generally used to compare different photodetector systems is the specific detectivity $D^*$ (in Jones = cm Hz$^{1/2}$ W$^{-1}$), which is *D* corrected for the detector optical area and bandwidth ($\Delta f$):

$$D^* = \frac{R_\lambda}{I_N}\sqrt{A_{opt}\Delta f} \quad (6)$$

*1.1.6 Time response and 3 dB bandwidth*

To quantify the time response $\tau_r$ of a photodetector under pulsed illumination the rise or fall time between 10% and 90% of the maximum response value are commonly used. The rise and fall times depend, on the one hand, on the photodetector electrical time constant $\tau_{RC} = RC$, where $R$ is the resistance and $C$ the capacitance viewed by the photodetector. On the other hand, the time response depends also on drift and diffusion processes. The drift component ($\tau_{drift}$) accounts for the movement of charge carriers traversing space charge regions, whereas the diffusion component ($\tau_{diff}$) describes the movement of charge carriers along charge-neutral regions. These components lead to

$$\tau_r = \sqrt{(2.2\tau_{RC})^2 + \tau_{drift}^2 + \tau_{diff}^2} \qquad (7)$$

for the time response (where the 2.2 factor arises due to the 10-90% definition of the response time). A detailed description of the response time of semiconductor photodiodes can be found in [18].

The 3 dB bandwidth ($BW_{3dB}$) is defined as the modulation frequency of the incident light when the responsivity decreases by 3 dB (= 0.707 times the low-frequency value). The bandwidth is approximately related to the response time of the detector by $\tau_r = 1/2\pi BW_{-3dB}$.

*1.2 Light coupling*

Extracting the responsivity of nanowire photodetectors from the measurements of the photocurrent as a function of the irradiance is not evident, mostly due to the non-obvious definition of the photodetector optical area [$A_{opt}$ in equation (1)] [6,19–21]. When a nanowire (or a nanowire ensemble) is exposed to an incident photon flux density $\Phi_{ph}$, precise calculation of the absorbed optical power, $P_{opt} = \Phi_{ph} A_{opt}$, requires understanding how a planar light wave interacts with the nanowires, which is a function of the wavelength, and the nanowire shape and refractive index. In the case of a nanowire ensemble or an array, the density and arrangement of the nanowires must also be taken into consideration.

Analytical investigations by Xu et al. [6] suggest that vertical nanowire arrays behave similarly to a concentrating lens or a parabolic mirror, due to the large refractive index contrast between the nanowires and the surrounding environment. This optical concentration effect, depicted in figure 1(a), results in an increase of the effectively absorbing cross-section. Following the analysis of Heiss et al. [19], we can define "absorption enhancement" as the ratio between the absorbed optical power and the optical power that impinges onto the geometric cross-section of the nanowire as

$$Absorption\ enhancement = \frac{A_{opt}}{A_{inc}} \qquad (5)$$

where $A_{inc}$ is the cross-section of the wire exposed to the incident light. In the case of a standing nanowire and a planar wave incident perpendicularly to the substrate, $A_{inc} = \pi r^2$, where $r$ is the nanowire radius (approximation of the nanowire by a cylinder with circular cross section). As an example to illustrate the relevance of the difference between $A_{opt}$ and $A_{inc}$, figure 1(b) describes the variation of the absorption enhancement as a function of the nanowire diameter ($2r$) and the wavelength of the incident light in a single GaAs nanowire standing on a silicon substrate. Note that the enhancement can reach a factor of 70, i.e. the optical area can be 70 times the geometrical area. Evidently, such considerations apply as well, with respective modifications, to nanowires lying horizontally on a substrate (or experiencing sideway illumination) [22,23].

Due to the complexity of the calculation of $A_{opt}$, responsivity values in the case of single nanowire photodetectors are often obtained using the geometrical cross-section, $A_{inc}$. In this case, the responsivity values can be significantly over-estimated.

In the case of nanowire ensembles, theoretical works have studied the dependence of the light absorption on the diameter of the nanowires, their geometry, and their geometrical arrangement on the substrate [21,24–27]. In particular, dense layers of randomly-spaced nanowires are known to enhance light absorption thanks to multiple scattering, which reduces the specular reflectance [28].

*1.3 Nanowire fabrication*

The fabrication of vertically-aligned semiconductor nanowires with an epitaxial relationship with the substrate has been demonstrated using a variety of techniques, such as molecular beam epitaxy, metalorganic vapor phase epitaxy, chemical vapor deposition, thermal evaporation, or hydrothermal growth. Independent of the growth technique, nanowires can be obtained by several methods: (i) catalytic (or metal-seeded) growth, (ii) catalyst-free growth, (iii) selective area growth, and (iv) top-down fabrication.

i) In the case of *catalytic growth* metal particles are dispersed on a substrate by means of metal evaporation and annealing (random distribution) or lithography, metallization and lift-off (position-controlled arrays). By increasing the temperature and introducing the reactants in the growth chamber, the metal particle forms an alloy with one or more reactants. When the alloy supersaturation is high enough, the growth of the semiconductor nanowire starts by precipitation at the particle/substrate interface, and continues layer by layer. This growth mechanism is often referred to as VLS (vapor-liquid-solid) [29], and it is maybe the most successful method in synthesizing single-crystalline nanowires. Figure 2 presents an schematic description of the method together with an example of growth of wurtzite GaAs nanowires using gold as a catalyst [30]. As a mayor limitation, the catalyst may contaminate the nanowire, altering its optical or electrical properties. The most commonly used metal is gold, but other materials have been considered. The catalyst must be a good solvent for the targeted nanowire material, ideally forming a eutectic compound.

When the catalyst is a constituent of the nanowire (e.g. Ga for the synthesis of GaAs nanowires), we speak about *self-catalyzed* or *self-seeded* growth [31,32]. The major advantage of a self-catalytic process is that it avoids contamination by foreign materials.

ii) Some materials do not require the presence of seed particles. The growth of nanowires happens spontaneously under certain growth conditions. In this situation, we speak about *catalyst-free growth* or *vapor-solid growth mechanism*. For example, GaN nanowires can be fabricated by plasma-assisted molecular beam epitaxy (PAMBE) under nitrogen-rich conditions [33–35]. Their strain-driven nucleation process is illustrated in figure 3. Likewise the synthesis of metal-oxide nanoribbons is possible by simply evaporating metal-oxide powders at high temperature [36].

iii) *Selective area growth* refers to a situation where the substrate is partially masked so that the material only grows in the mask openings [37–41]. By tuning the growth conditions, nucleation is obtained only in the areas where the substrate is exposed. Growth can be highly anisotropic, so that the material grows in one direction only keeping the cross section defined by the mask. As an example, figure 4 shows GaN nanowires generated by selective area growth on a GaN-on-Si(111) structure using a thin (5 nm) e-beam patterned Ti mask [42]. Growth was performed by PAMBE. In general, in a selective area growth process, the growth rate depends on the distance between mask apertures.

The selective area growth reduces the variations in nanowire size and length that are characteristic of self-organized processes. As a possible drawback, good selectivity between the mask and the nucleation sites requires relatively high temperature, which might in certain cases complicate the growth of heterostructures or ternary/quaternary alloys. Another issue to be kept in mind is that the nanowire length depends not only on the growth time and precursor fluxes, but also on the distance between mask apertures.

iv) Finally, it is also possible to fabricate nanowires following a *top-down* approach, i.e. etching down an originally planar structure [8,43,44]. The etching pattern can be defined using electron lithography, nanoimprint or colloidal masking, for instance. Top-down nanowires present better wire-to-wire homogeneity (even compared to selective area growth), and well controlled doping profile and heterostructure dimensions. This facilitates their integration into large-scale devices. However, in this process the critical thicknesses are determined by the two-dimensional growth, and the dislocations generated in the two-dimensional epitaxial growth remain in the patterned nanowire array. There is also a risk of structural damage during the plasma etching process, which might lead to a degradation of the optical properties. In some cases, damaged areas can be removed by wet etching, recovering the optical performance.

Figure 5 presents an example of GaN nanowires obtained by a *top-down* approach [45]. In this case, the etch mask consisted of a monolayer of 3 µm diameter silica colloids that was self-assembled on the GaN surface in a Langmuir-Blodgett trough. GaN posts are then dry etched, which results in tapered rods. Finally, the etch-damaged areas are removed and the rods are thinned by anisotropic wet etching.

*1.4 Structural properties*

This review paper focusses on nanowire photodetectors crystalizing in wurtzite structure. Figure 6(a) depicts the wurtzite lattice, which shows hexagonal symmetry. In such a lattice, directions and planes are described using four Miller-Bravais indices (hkil), where l is associated to the vertical axis of the hexagonal prism, and h, k, and i = −(h + k) are associated to vectors contained in the base of the prism, pointing to vertices of the hexagon that are separated by 120°. The c-(0001), m-(10-10) and a-(11-20) planes are outlined in figure 6(a). The [0001] axis is considered positive when the vectors along the bonds between the group II or III atoms and the group VI or V atoms along <0001> points from metal to the group VI or V. Conventionally, (0001) crystals are called metal-polar and (000-1) crystal are called X-polar, where X is the group VI or V element.

The wurtzite unit cell is non-centrosymmetric, meaning it lacks inversion symmetry, which results in materials with significant pyroelectric and piezoelectric properties. The difference in electronegativity between the two atomic species that constitute the lattice results in a displacement of the electrons of the bonds towards the more electronegative atom. The electric dipoles associated to the bonds are not mutually compensated due to the lack of symmetry of the crystal, which leads to a macroscopic spontaneous polarization field along the <0001> axis. The value of such polarization depends on the ideality of the crystal, the anion-cation bond length, and the electronegativity of the atoms involved. Therefore, it varies from one material to another (see table 1). Bringing two of these materials into contact leads to a charge sheet at the interface. These spontaneously formed charge sheets significantly modify the band structure across the interface, which can be used in order to band engineer specific heterostructures. An example can be seen in figure 6(b), which presents the band profile of a GaN/AlN quantum well.

An additional polarization field can appear as a result of the lattice deformation due to misfit stress. Stress leads to a deformation of the crystal described by Hooke's law:

$$\sigma_{ij} = \sum_{kl} c_{ijkl}\varepsilon_{kl} \tag{8}$$

where $c_{ijkl}$ is the elastic tensor, and $\sigma_{ij}$ and $\varepsilon_{kl}$ represent the stress and strain respectively. The presence of strain induces a polarization vector whose components are given by

$$P_j = \sum_{kl} e_{jkl}\varepsilon_{kl} \tag{9}$$

The terms $c_{ijkl}$, and $e_{jkl}$ are transformed to $c_{mn}$, $e_{jm}$ by replacing $m,n$ = {*xx, yy, zz, yz, zx, xy*} with $m,n$ = {1, 2, 3, 4, 5, 6}. Due to the crystallographic symmetry of the wurtzite structure, the only non-zero elastic constants are $c_{11}$ = $c_{22}$, $c_{12}$ = $c_{21}$, $c_{13}$ = $c_{31}$ = $c_{23}$ = $c_{32}$, $c_{33}$, $c_{44}$ = $c_{55}$, $c_{66}$ = $(c_{11} - c_{21})/2$, and the non-zero piezoelectric constants are $e_{31}$ = $e_{32}$, $e_{33}$, and $e_{15}$ = $e_{24}$.

Among the compounds that generally take the wurtzite structure are III-nitrides (GaN, AlN, InN), II-oxides (ZnO, ZnMgO, ZnCdO), II-sulfides (ZnS, CdS) and II-selenides (ZnSe, CdSe). In the conditions required for growth in the nanowire geometry, some semiconductors that normally crystallize in the zinc-blende or diamond configuration can also take the wurtzite structure (III-arsenides, III-phosphides, Si). However, in these latter cases, it is difficult to obtain a pure wurtzite structure, and wires are often characterized by the presence of stacking faults, twin defects, and polytypisms [46]. Such structural defects can induce scattering of charge carriers and phonons, and modify the energy band gap in an uncontrolled manner, with potentially detrimental effects on the device performance.

The values of spontaneous polarization, elastic constants and piezoelectric constants of various wurtzite materials are presented in table 1, together with their lattice constants and energy band gaps.

Looking at their heterostructuring capabilities the GaN and ZnO material families deserve particular attention. In the case of III-nitride semiconductors, the band gap of the ternary alloys is obtained by a quadratic interpolation of the energy of the corresponding binary compounds $[E_g(x) = (1-x)E_g^A + xE_g^B - bx(1-x)]$, using bowing parameters $b$ = 0.68 eV for AlGaN, $b$ = 2.1 eV for InGaN and $b$ = 4.4 eV for AlInN [47]. For wurtzite MgZnO and CdZnO, bowing parameters in the range of 0.48-0.87 eV and 0.023-1.30 eV have been reported [48,49]. The conduction band offsets between binary compounds are represented in figure 6(c) (for nitrides, ≈ 1.8 eV for AlN/GaN [50] and ≈ 2.2 eV for GaN/InN [51], and for oxides, the conduction band offsets are calculated from the valence band offsets in ref. [52]: ≈ 1.2 eV for MgO/ZnO and ≈ 0.1 eV for ZnO/CdO). In view of the application of nanowires in solar cells, there is also an interest in the fabrication of type-II heterostructures, as an alternative to p-n junctions. Figure 6(d) presents recent calculations of the band offsets at ZnO/ZnX (X = S, Se, and Te) heterointefaces [53]. Note that in this case, zinc-blende ZnS, ZnSe and ZnTe are considered. There is a considerable dispersion of these offsets in the literature (see also refs. [54–56], for instance).

Self-assembled GaN nanowires synthesized on sapphire substrates by metalorganic vapor phase epitaxy (MOVPE) grow along the <0001> polar axis and they present generally {1-100} sidewall facets and mixed polarity [57–59], with predominance of N polarity but with inversion domains contained in the wires. The predominance of N or Ga polarity depends on the substrate, substrate treatment and growth conditions. Thus, pencil-like Ga-polar nanowires can be grown on non-nitridated sapphire under low precursor flows [60]. In the case of nanowires obtained by selective area growth on sapphire, the core of the wires is generally N polar, and it is sometimes surrounded by a Ga-polar shell that nucleates on the opening rim [58]. Predominant Ga polarity is obtained by selective area grown Si(111) substrates using AlN as nucleation layer [61].

In the case of GaN nanowires grown by PAMBE, self-assembled nanowires grow generally along the [000-1] axis and present {1-100} sidewall facets. This is independent of the substrate (silicon, sapphire, graphene, SiC, …). It is also relatively common that the nanowires contain inversion domains [62]. The more reliable method to obtaining Ga-polar nanowires by PAMBE is the selective area growth on Ga-polar GaN substrates using metallic or dielectric masks [63], as illustrated in figure 4.

In summary, GaN nanowires grow along the polar axis and present nonpolar, generally m-plane facets. This implies that axial heterostructures are affected by the polarization-related internal electric field, but this is not the case for core-shell heterostructures.

In contrast, in the case of ZnO, there are three families of fast-growth directions, namely <0001>, <1-100> and <11-20>, where the growth axis depends on the relative surface energies and kinetic factors. Typically, the terms "nanowire" or "nanorod" refer to hexagonal prisms that grow along <0001> and present <1-100> or <11-20> facets, and "nanobelt" refer to rectangular prims that can grow along <0001>, <1-100> or even <11-20> and present <11-20>, <1-100> or <0001> facets [64].

Thermal sublimation techniques allow the controlled growth of all these nanostructures [64], depending on the deposition temperature and use of catalysts. However, nanowires grown by molecular beam epitaxy (MBE) [65,66] or MOVPE [67] grow typically along [0001] (Zn-polar direction). The [000-1] growth direction can be obtained in the case of catalyst-assisted (Au) growth by metalorganic vapor phase epitaxy [66], as illustrated in figure 7.

## 2. Nanowire metal-semiconductor-metal photodetectors

Metal-semiconductor-metal photodetectors consist of a semiconducting material sandwiched between two metal contacts, which can be ohmic or Schottky (rectifying) [68]. In general, a device with two ohmic contacts behaves as a photoconductor, i.e. a resistor whose ohmic value depends on the impinging light intensity. In the case of planar structures, photoconductors are slow (seconds) devices that operate under bias and can display extremely high values of responsivity, but their response is strongly sublinear with the optical power and they present problems of spectral selectivity since they are very sensitive to absorption below the semiconductor band gap [68].

Replacement of one of the contacts by a Schottky contact results in the presence of a depletion region below the Schottky metallization which favors the collection of photogenerated carriers by drift transport. Typically, such devices react linearly to the incident irradiation, they do not present internal gain, and their speed is limited by resistance × capacitance (RC) factors. Schottky diodes can operate at zero bias, although applying negative bias on the Schottky contact increases the length of the depletion region resulting in higher responsivity and faster response (lower capacitance).

If the two contacts are Schottky barriers, a situation of total depletion between the contacts may occur. In such a situation, the devices present space-charge-limited dark current and photogenerated carriers drift fast between the contacts. Generally, the photocurrent scales linearly with the optical power. Photodetecting requires external bias, and gain might appear if the devices are operated at very high voltages.

Once introduced the different kinds of metal-semiconductor-metal detectors that can be implemented in planar geometries, we discuss below the results when such devices are fabricated using nanowires. Single nanowire photodetectors can be fabricated by dispersion of nanowires on an insulating substrate, followed by contacting using typically electron beam lithography, with the result described in the schematic in figure 8(a). On the other hand, ensemble nanowire photodetectors consist of a layer of standing nanowires which is planarized before contacting, as described in figure 8(b). There are a number of publications on single nanowire photoconductors using a variety of wurtzite materials (GaN [69,70], AlN [71], InN [72], ZnO [73–76], ZnSe [77], ZnS [78–80], CdS [81,82], InP [83]). Here we summarize their main characteristics.

## 2.1 Linearity and responsivity

To identify the role of the nanowire on the photodetector performance, we will in the following focus on the properties of single nanowires. In the case of single nanowire photoconductors, the variation of the photocurrent as a function of the incident optical is often sublinear [70,73,74,84–88], as illustrated in figure 9(b) for PAMBE-grown n-i-n (400 nm n-type Si-doped / 400 nm undoped / 400 nm n-type Si-doped) GaN nanowires with the characteristics and measuring setup described in figure 9(a) [70], or in figure 10 for 10-15-µm-long non-intentionally-doped ZnO nanowires grown by chemical vapor deposition [73]. Experimental results are often described with the relation

$$I_{ph} \propto P_{opt}^{\beta}$$

where the degree of linearity can be quantified with the exponent $\beta$. In the case of $\beta = 1$ the detected photocurrent is linear, whereas $\beta < 1$ describes a sublinear dependence. In the case of pulsed photocurrent measurements using a synchronous detecting system (light chopper and lock-in amplifier), increasing the chopping frequency leads to an improvement of the linearity (higher $\beta$) [70,89], as illustrated in figure 9(b). However, in the case of GaN nanowires, it was demonstrated that the chopping frequency does not have any effect on the spectral response of the photodetector [see figure 9(c)], in contrast to similar measurements on planar photoconductors [68].

The nonlinearity of nanowire photodetectors, observed even in nanowires that do not contain dislocations, stacking faults or inversion domains, suggests that the photoresponse is dominated by the redistribution of charge at the surface levels [5,70,74,90]. The surface states (mostly hole traps, with the exception of materials such as InN or InAs) determine the location of the Fermi level at the nanowire sidewalls, and generate a depletion region that propagates into the nanowire core [see figure 11(a)]. Illumination is known to modify the Fermi level pinning at the sidewalls [91], which results in a decrease of the depletion region depth and the corresponding enlargement of the diameter of the conductive core. This phenomena is sometimes called photogating effect [88,92], and it justifies the nonlinearity of the photocurrent via the second term of eq. (4) [70,93]. An additional contribution to the nonlinearity comes from an increase of the lifetime of photogenerated electrons due to the trapping of holes at surface states [5,69,94].

There is a correlation between the degree of linearity, the nanowire diameter, and the dark current [90,93]. Statistically, the nanowires become linear ($\beta \approx 1$) and highly resistive when the diameter is smaller than a certain threshold. The dark current can drop several orders of magnitude (nA to µA), as shown in the inset of figure 12(a) for GaN nanowires with a diameter smaller than $\approx$ 80 nm. When plotting the value of $\beta$ as a function of the dark current [figure 12(a)], nanowires with low dark current (small diameter) are almost linear, whereas those with high dark current (large diameter) are systematically sublinear. Figure 12(b) shows that the spectral response of the wires is the expected response of GaN, independent of the nanowire diameter, which means that the photocurrent originates in the nanowires and the influence of the carrier wafer ($Si_3N_4$-on-silicon in this case) is negligible.

These differences in linearity can be understood looking at the potential profile along the nanowire diameter, as shown in figures 11(a-c). As it was described above, the Fermi level pinning at the nanowire sidewalls generates a depletion region that propagates into the nanowire core. In large nanowires [figure 11(a)], the depletion region is a shell that envelops the nanowire conductive core. However, in narrow-enough nanowires [figure 11(b)], the surface states can fully deplete the nanowires, which explains a drastic drop of the dark current for diameter below a critical value of the diameter [5,69,93,95].

Illumination modifies the location of the Fermi level at the surface [91] by changing the charge distribution in the surface states or by interaction with adsorbates, especially oxygen [73,88,91]. As a result, in large, partially-depleted nanowires, light induces a modulation of the diameter of the conductive core. Looking back to eq. (4), this means that the photoinduced variation of the conductance, $\Delta G$, can be dominated by the variation of the conductive cross-section, $\Delta A$, which explains the nonlinearity of the response. On the other hand, the band bending induces a radial separation of the photogenerated electrons and holes. The holes are attracted towards the sidewalls, and the electrons, in the core, need to cross a radial potential barrier to recombine, which results in an enhancement of the carrier lifetime.

In the case of very narrow, fully-depleted nanowires [figure 11(c)], the incident light is not able to generate a conductive core. Furthermore, the effect of surface traps on the carrier lifetime is mitigated because the potential barrier between the center of the wire and the surface is lower than in partially-depleted structures. Therefore, it is possible to reach a situation where the responsivity is dominated by the first term of eq. (4), i.e. ΔG scales linearly with the optical power, as it is proportional to the amount of photogenerated carriers, Δn. Of course, saturation can be reached at very high laser pumping densities, due to the reduction of the depletion region depth induced by the very high density of photogenerated carriers.

In this discussion, we have considered that the Fermi level pinning at the nanowire sidewalls originates an upwards band bending, hence pushing electrons towards the core of the wire. This is the case in nanowires based on GaN, GaAs, ZnO, ZnS, CdS, etc. However, it is known that the use of materials like InAs [96–98] or InN [99–101] results in the opposite configuration: the Fermi level is pinned deep into the conduction band, and this originates an accumulation of electrons at the sidewalls.

There is a general agreement that nanowire photodetectors present high responsivity. However, the precise value of reponsivity can depend on the optical power and on the light modulation frequency. Therefore, it is important to indicate the degree of linearity of the detector, and the incident optical power for which the responsivity is measured. Furthermore, it is also imperative to indicate the optical area, $A_{opt}$, which is taken into account in the calculation (see discussion in section 1.2), to transmit a clear idea of the meaning of the responsivity value and allow comparative studies. Some authors prefer reporting gain instead of responsivity. Let us remind that gain and responsivity are linearly linked by eq. (2), and the quantum efficiency ($\eta$, number of photogenerated electrons per incident photon) is generally assumed to be unity. Keeping these considerations in mind, experimental values of gain in the range of $10^5$-$10^8$ have been reported in single nanowire photodetectors based on ZnO [73,74], ZnS [79,102,103], ZnSe [77], GaN [70,87,104], InN [72], GaAs [86,105], or InAs [106,107]. In general, these high gain values can be attributed to the effect of light on the surface depletion region (i.e. on the conductive are of the device) and to the enhancement of the carrier lifetime due to the radial separation of photogenerated electrons and holes.

*2.2 Time response*

As described above, the radial separation of photogenerated electron and holes due to the surface band bending introduces a delay in the time response of nanowire photodetectors. When the light is switched off, the excess electrons, in the nanowire conductive core, have to overcome a potential barrier to recombine with the holes, located at the sidewalls. During the recombination process, the barrier height depends on the remaining carrier excess, so that the photocurrent decay can be strongly nonexponential [108]. As a result, the measured time response is a function of the incident optical power, with the response being faster for higher optical excitation. On the other hand, fully depleted nanowires are responding faster than partially depleted wires [95], since the potential barrier for electron-hole recombination is lower in the former case.

It is important to note that the time response of nanowire metal-semiconductor-metal photodetecotrs (in the second to picosecond range, depending on the device architecture) is significantly faster than the persistent photoconductivity typically observed in planar photoresistors, where the photocurrent transients can be in the range of hours [68]. It has been argued that the presence of deep defects such as oxygen vacancies in ZnO or nitrogen vacancies in GaN might play a role on the presence of persistent effects in nanowires [5,73], but it is difficult to assess the relevance of such defects in comparison with undeniable role of the surfaces and adsorbates. On the contrary the excellent spectral selectivity of GaN and ZnO nanowire photoconductors [70,73] testifies to the low point defect density in the nanowires.

*2.3 Effect of the environment*

As discussed above, the chemical environment can modify the energy location of the Fermi level at the nanowire sidewalls, with the subsequent effect on the dark current, responsivity/gain and time response. Therefore, these devices are particularly suitable for gas and chemical sensing. As an example, a comparison of the behavior of GaN, ZnO or InAs nanowire photodetectors in vacuum and in the air shows that the vacuum environment leads to higher dark current, higher responsivity and much slower photocurrent transient [73,95,106,109–111]. In ZnO and GaN, all these trends are consistent with the Fermi level at the surface getting deeper into the gap with respect to its location in the air (in figure 11, higher $\Delta E$ in vacuum than in the air) due to the desorption of highly-electronegative adsorbed oxygen induced by UV illumination [73,88,91]. The sensitivity to the environment can be attenuated by inserting heavily doped regions in the nanowire [109], or by dielectric passivation of the surface [106].

*2.4 Nature of the contacts*

Metal-semiconductor-metal nanowire photodetectors with two ohmic contacts require bias, and they have high responsivity, often at the price of a slow response time. If one of the contacts is replaced by a Schottky contact, the dark current drops drastically as a result of the potential barrier at the metal-semiconductor interface. Under illumination, photogenerated electrons and holes are axially separated due to the electric field in the proximity of the Schottky contact. This allows operation at zero bias, which has motivated some authors to refer to these devices as "self-powered devices" [112]. The response time is shortened since transport under illumination is dominated by the drift of minority carriers towards the Schottky contact, and, in the absence of light, the current is quenched by the presence of the Schottky barrier. These processes have been demonstrated, for instance in ZnO nanowires [113–116]. In contrast to planar Schottky photodiodes, devices based on nanowires remain generally sublinear and with a millisecond time response, which is an indication of the relevance of surface phenomena in the process of carrier collection.

When both contacts are replaced by Schottky diodes, devices are generally referred as metal-semiconductor-metal photodiodes. In planar structures, the back-to-back Schottky configuration is interesting to obtain total depletion between contacts. However, in the nanowire geometry, it is easier to obtain total depletion radially, via the sidewall surfaces, than axially, where very low residual doping levels would be required to get a Schottky depletion length comparable to the nanowire length. The use of two Schottkky contacts results in a drastic decrease of the dark current, but the nominal symmetry of the device prevents zero-bias operation.

## 3. Nanowire p-n junction photodiodes

In p-n photodiodes, electrons and holes photogenerated near the junction drift in opposite directions due to the internal electric field in the space charge region. Similarly to Schottky photodiodes, p-n photodiodes can be operated at zero-bias, but speed and responsivity are enhanced when the junction is reverse biased, owing to the increase of the space charge region length. These devices are expected to display lower dark current than metal-semiconductor-metal photodetectors, since the junction behaves as a potential barrier that inhibits carrier transport in the dark. Furthermore, the capacitance associated to the depletion region of the p-n junction can be easily smaller than that of a Schottky photodiode, since the p-n potential barrier is approximately the semiconductor band gap, generally larger than a Schottky barrier. Under zero-bias or reverse bias, p-n photodiodes react linearly to the incident optical power and present zero gain. Gain associated to impact-ionization phenomena (avalanche amplification) can appear when the devices are strongly reverse biased (several times the semiconductor band gap). Gain can also occur under forward bias or in the case of reverse-biased junctions with important leakage current, since in these cases the response can be dominated by the photoconductor-like behavior of the conductive p and n type regions.

Nanowire-based p-n photodiodes have been reported using a number of materials. In nanowire devices where both p and n regions are contained in the same nanowire, the arrangement can be radial [core-shell junction, as described in figure 13(a), e.g. GaAs [7,117], GaAsP [118,119] or GaN [120]] or axial [as described in figure 13(b), e.g. GaN [121,122], GaAs [123], or ZnO [124]]. The p-n junction can be implemented as a homojunction (both p- and n-doped zones are of the same material) or a heterojunction (p- and n-doped zones combine different materials). Homojunctions have the obvious advantage of avoiding any lattice mismatch, and therewith losses due to carrier recombination at structural defects at the junction. However, the doping levels can be limited by material or growth issues, e.g. it is difficult to obtain p-type ZnO or uniformly doped p-type GaN nanostructures. As alternative solutions, p-n photodiodes can be obtained by crossing two nanowires [figure 13(c)], one of them n-type and the other one p-type [77,116,125–129], branching [130–132] [figure 13(d)], or growing n(p)-type nanowires on a p(n)-type substrate [figure 13(e), e.g. n-i-InP nanowires on p-InP substrate [133], n-ZnO nanowires on p-GaN substrate [134,135], p-ZnO nanowires on n-ZnO substrates [136] or p-InGaAs nanowires on n-GaAs substrate [137]].

In the case of compound-semiconductor nanowires, avalanche photodiodes have been demonstrated using a crossed n-CdS/p-Si heterojunction [125], an InP axial p-n junctions incorporating an InAsP quantum dot [138] or an InAsP section ($n^+$-InP/n-InAsP/n-InP/$p^+$-InP, illustrated in figure 14) [139], radial GaAs p-n junctions [140,141], and p-i-InGaAs nanowires on n-GaAs substrates [142]. These devices are interesting for application in single-photon counting. For this purpose, semiconductor nanowires offer advantages such as enhanced absorption due to optical resonance effects, elastic relaxation of the misfit strain in heterostructures, and compatibility with silicon technology. An additional advantage that has been recently proposed is associated to the structure of a nanowire ensemble photodetector. Among the total amount of nanowires constituting one photodiode, each avalanche event is confined in a single nanowire, which means that the avalanche volume and the number of filled traps can be drastically reduced. This leads to an extremely small afterpulsing probability compared to conventional single-photon avalanche photodetectors (SPADs). Farrell et al. have recently demonstrated nanowire-ensemble SPADs with a dark count rate below 10 Hz, due to reduced fill factor, with photon count rates of 7.8 MHz and timing jitter less than 113 ps [142].

## 4. Heterostructured nanowire photodetectors

A heterostructure sequences a number of materials in a specific order to engineer the energy bands for certain purposes, e.g. charge carrier separation, quantum confinement, current blocking or tunneling. There are two ways of heterostructuring a nanowire. One is a radial sequencing of the materials which results in so called core-shell structures, as described in figure 15(a). The other is an axial sequencing of the materials which results in so called axial or dot-in-a-wire heterostructures, illustrated in figure 15(b).

*4.1 Radial (core-shell) nanowire photodetectors*

The fabrication of core-shell nanowire heterostructures can obey several motivations:

(i) To provide a *photoconductive nanowire with a passivating envelop* that improves its stability and reduces the effect of surface phenomena (e.g. surface recombination or chemical sensitivity). Examples of this application are GaAs/AlGaAs [143] or ZnO/ZrO$_2$ [144] core/shell structures.

(ii) To implement a *core-shell p-n junction photodetector*. In this case, the radial heterojunction is considered advantageous with respect to an axial heterojunction since the direction of light incidence is normal to the direction of charge separation. Therefore, this geometry provides short collection lengths (determined by the nanowire diameter, which is smaller than the carrier diffusion length) compatible with a large absorbing volume (determined by the nanowire length, which can be much longer than the absorption depth) [145]. As a result, the carrier collection efficiency can be improved without reducing the total absorption. However, in view of its application in solar cells the radial heterojunction geometry complicates the implementation of tandem architectures considerably [146,147].

(iii) To implement a *type-II heterostructure* which separates electrons and holes radially, thus increasing the carrier lifetime. Such type-II heterostructures have been proposed as an alternative to p-n junctions for the fabrication of solar cells [148]. Examples of type-II core/shell nanowire heterostructures are ZnO/ZnTe [149–151], ZnO/ZnSe [152–154], ZnO/ZnS [155,156] or ZnO/CdTe [157–160]. Note that the most stable crystallographic structure of ZnTe, ZnSe and ZnS is cubic zinc blende. As a result, an interesting feature in these kind of heterostructures is that the shell can present a different crystallographic structure than the wurtzite ZnO core. There is a significant dispersion in the values of band offsets in the literature [53–56]. Figure 6(d) represents a recent set of values taken from ref. [53].

*4.2 Axial nanowire photodetectors*

The incorporation of axial heterostructures in nanowire photodetectors may serve various purposes, e.g. designing quantum wells to tune the absorption spectrum or introducing a band bending leading to charge carrier separation. Within quantum wells, nanowire photodetectors can make use of either band-to-band transitions, whose energy depends on the semiconductor bandgap, or intersubband (ISB) transitions, i.e. transitions between confined electron states located within the conduction band or within the valence band. An ISB transition can occur only when there is an excess of electrons (if the transition involves conduction band states) or holes (if the transition involves valence band states). The ISB transition energy is determined by the size of the nanostructures and is independent of the semiconductor band gap.

*4.2.1 Interband nanowire photodetectors*

In the near-infrared spectral range, nanowire photodetectors incorporating axial heterostructures have been demonstrated using the InAs/InAsP [161] and InAsP/InP [138,162,163] material systems. As an outstanding result, InP nanowire photodetectors with a single InAsP well as light absorbing element demonstrated a quantum efficiency of 4% [162]. Under resonant excitation, the photocurrent scales linearly with the optical power and it is enhanced for light polarized parallel to the nanowire axis. These first results were promising for the development of photodetectors with subwavelength special resolution. In a second stage, a single InAsP well displaying photocurrent in the 963-1007 nm range was located in the avalanche multiplication region of a nanowire photodiode, which allowed the demonstration of single photon detection with a nanowire photodiode [138].

In the visible domain, a GaAs/GaP multi-quantum-well structure was used to tune the photocurrent spectral response of a GaP nanowire photodetector [164]. The presence of 15 periods of GaAs/GaP (13 nm/ 15 nm) inserted in the i-region of a GaP p-i-n junction nanowire did not have a significant impact on the electrical behavior of the structure, but it red shifted the photocurrent spectral cutoff from 500 nm for pure GaP to 550 nm. In this case, an AlP shell was used to envelop the whole nanowire as a passivation method.

In the ultraviolet spectral range, selective photodetection is generally implemented using ZnO or GaN based nanowires. These materials present strong polarization along the <0001> crystallographic axis, which is generally the growth axis of the nanowires. At an axial heterointerface with a material of different polarization, the local charge distribution does not get compensated, which results in a fixed charge sheet located at the heterointerface. Such polarization-induced charge sheets modify significantly the band structure, e.g. generating an internal electric field in the quantum wells (see figure 6(b)], which leads to the separation of electron and hole wavefunctions. Therefore, the oscillator strength of the band-to-band optical transition decreases, and the transition energy decreases too. This is known as quantum confined Stark effect. Taking the polarization differences into consideration is hence critical to understand the performance of axial heterostructures.

The effect of inserting a 3.6-nm-thick AlN section into an n-i-n GaN single nanowire photodetector was studied by den Hertog et al. [110,165], with the results illustrated in figure 16. The nanowires were synthesized catalyst-free by PAMBE under N-rich conditions. Structurally, the AlN insertion resulted also in the formation of an AlN shell that enveloped the GaN base of the nanowire. This thin ($\approx$ 0.5 nm) AlN shell was in turn covered by a GaN shell with a thickness of 4-10 nm. These shells are due to the AlN and GaN radial deposition during growth. From the electrical viewpoint, the presence of the AlN insertion blocks the electron flow through the GaN core forcing a drop of the dark current by at least 3 orders of magnitude. The polarization difference between AlN and GaN results in an asymmetric potential profile that manifest in rectifying current-voltage characteristics in the dark, and zero-bias photoresponse under illumination. The insertion of the AlN barrier in the n-i-n GaN nanowire does not degrade the spectral selectivity of the structures. However, the GaN/AlN/GaN core/shell structure provides a parallel conduction path, i.e. the current can flow through the GaN outer shell, close to the nanowire sidewalls, as indicated in figure 16(a). This surface conduction path increases the sensitivity of the photocurrent to the environment and in particular to the presence of oxygen [see figure 16(d)], which can be interesting for the fabrication of chemical sensors, but it is undesirable in GaN/AlN photodetectors since it reduces the environmental stability and masks the advantages of the inserted heterostructure.

The presence the GaN outer shell can be avoided by increasing the growth temperature. Axially heterostructured single nanowire photodetectors containing a wurtzite GaN/AlN superlattice without outer GaN shell were first demonstrated by Rigutti et al. [166] in 2010. PAMBE-grown nanowires incorporating a stack of 20 AlN/GaN (2-3 nm/ 1-5 nm) quantum wells were considered. In comparison to nanowires without any heterostructure, the insertion of the superlattice resulted in a reduction of the dark current that could reach 8 orders of magnitude (in the case of nanowires without external GaN shell). The presence of the superlattice resulted also in an enhancement of the so-called photosensitivity factor (photocurrent divided by the dark current); however, the responsivity dropped significantly, going from 8-40×10$^4$ A/W to 100-2000 A/W at -1 V bias in nanowires with the GaN/AlN superlattice (values corresponding to illumination with 5 mW/cm$^2$ of UV light at λ = 300 nm). Note that the responsivity was calculated assuming that the optical area is equal to the geometrical area of the nanowire. The spectral response is dominated by the GaN sections of the nanowire, but the spectral contribution from the GaN wells, located in the 450-400 nm spectral range, can be resolved in photocurrent spectroscopy measurements performed at room temperature. The photoresponse drops significantly with decreasing temperature, which demonstrate that photogenerated carrier extraction is thermally assisted.

Using a GaN/AlN (2.7 nm/ 5.3 nm) superlattice without GaN outer shell (see figure 17), Lähnemann et al. [89] showed a photocurrent that scaled sublinearly with the optical power. The spectral response clearly shows a sensitivity to wavelengths below 360–380 nm (cut-off associated to transitions to the first excited level in the quantum wells), and a rejection of longer wavelengths. Persistent photoconductivity is observed, with an initial decay time on the order of a few milliseconds, but it includes also slower components. Statistical measurements of dispersed nanowires from the same wafer present a significant dispersion in the values of dark current (10$^{-12}$ to 10$^{-6}$ at +1 V bias) and photocurrent (10$^{-9}$ to 10$^{-2}$ A). This observation highlights the fact that conclusions should not be based on observations of only one or two nanowires from a given sample. The dispersion is explained as due to the coalescence of nanowires with displaced heterostructures, reducing the effective length of the heterostructure. As illustrated in figure 17(d), lower dark currents correlate with higher photosensitivity factors, which is explained by a larger number of nanodisks contributing to the photocurrent.

Also using PAMBE-grown GaN nanowires, Spies et al. [167] demonstrated an improved design which takes advantage of the polarization-induced internal electric field to improve the responsivity. The structure is described in figure 18(a). In the GaN/AlN superlattice, the width of the GaN wells was reduced to 2.5 nm to facilitate the carrier extraction. Additionally, an asymmetry was introduced in the doping profile: The stem and the superlattice remained undoped, whereas the cap was heavily doped n-type. This should avoid the pinning of the Fermi level at the conduction band in the stem, and enlarge the regions with internal electric field in the structure. With this new architecture, responsivity values of 150±30 kA/W at +1 V were demonstrated, which is roughly 2-3 orders of magnitude larger than in previous reports [166]. Furthermore, depending on the application of positive or negative bias, there is an enhancement of the collection of photogenerated carriers from either the GaN/AlN quantum wells of the superlattice (in the ≈ 280-330 nm ultraviolet B range) or from the GaN stem region (in the ≈ 330-360 nm ultraviolet A range), respectively [see figure 18(b)]. This behavior can be understood by analysis of the potential profile along the nanowire growth axis, illustrated in figures 18(c-e). At zero bias, the band profile presents a triangular shape induced by the polarization-induced charge sheets at the interfaces between the GaN stem and the superlattice. These fixed charges create a large depletion region in the GaN stem and an internal electric field along the whole superlattice, which is almost fully depleted. On the contrary, at the interface between the superlattice and the GaN cap the difference in polarization pins the Fermi level at the conduction band. The fact that the cap is heavily doped leads to flat bands all along the cap section. In summary, the sample contains two areas with internal electric field: the superlattice with electric field pointing toward [0001] and the depletion region in the stem with electric field pointing toward [000-1]. The application of bias determines the dominant electric field. Under negative bias [figure 18(c)], the bands flatten along the superlattice, and photogenerated electrons absorbed in the GaN stem can be collected. Under positive bias [figure 18(e)], the depletion region in the stem shrinks and the collection of carriers photogenerated in the superlattice is favored.

*4.2.2 Intersubband nanowire photodetectors*

Intersubband (ISB) devices rely on transitions between quantum-confined electron levels within the conduction band (or within the valence band) of semiconductor heterostructures. As the energy associated to such transitions is relatively small, the operation wavelengths of ISB devices are in the infrared (IR) spectral region, in the 1–50 µm range. ISB transitions are governed by a polarization selection rule: to interact, the incident radiation needs to have an electric field in parallel to the direction of confinement in the heterostructure [transverse magnetic (TM) polarized light], i.e. there is no interaction with transverse-electric (TE) polarized light. A comprehensive introduction to ISB physics in quantum wells can be found in the works of Bastard [168] or Liu and Capasso [169]. Quantum cascade lasers [170] and quantum well infrared photodetectors (QWIPs) [171] are well-known illustrations of ISB devices.

In the 5–30 µm wavelength window, III-As based QWIPs can achieve picosecond response times, outperforming interband devices in terms of speed. GaN and ZnO materials open perspectives for room-temperature operation of ISB devices both in the near-IR (1–5 µm) and in the mid- to far-IR (5–30 µm) spectral ranges [172,173]. Even more importantly, they offer the opportunity to integrate nanowires as active media, which represents the ultimate downscaling of devices incorporating quantum well superlattices.

Following the trend of planar technologies, the logical choice of materials for nanowire ISB devices would be III-arsenides. In this line, the observation of resonant tunneling transport through quantum confined levels in an InP/InAs double-barrier nanowire heterostructure was already reported in 2002 [174]. The current-voltage characteristics displayed negative differential resistance (NDR) with peak-to-valley ratios of up to 50:1 and current densities of 1 nA/µm$^2$ at low temperatures. However, the pronounced polytipism in III-As nanowires has so far hindered the observation of ISB transitions in bottom-up GaAs-based heterostructured nanowires (self-assembled or selectively nucleated).

In spite of these limitations, broadband ISB photodetectors were fabricated by planarization and contacting of an InP nanowire array containing a InAs$_{0.55}$P$_{0.45}$ quantum wells [175–177]. The arrays were synthesized by selective area growth by MOVPE and planarization was performed by spin coating with photoresist S1818. The infrared response covered the 3-20 µm spectral range and was observed under normal incidence excitation. This was explained by the excitation of the longitudinal component of optical modes in the photonic crystal formed by the nanowire ensemble, combined with the non-symmetric potential profile of the wells, which are spontaneously formed during growth.

In the case of the wurtzite III-nitride material system, the lower density of structural defects in self-assembled nanowire heterostructures has made it possible to observe ISB transitions the short- and mid-wavelength IR ranges [178–182]. Figure 19 presents the structural properties of a Ge-doped GaN/AlN nanowire heterostructure that displays ISB absorption around 1.5 µm, with the θ–2θ x-ray diffractogram being compared with that of a similar Si-doped structure, and those of planar layers with the same periodicity. In spite of the wire-to-wire inhomogeneities and the variations of tilt and twist in the nanowire ensemble, several satellites of the superlattice reflections are clearly resolved.

Figure 20 compares the ISB absorption observed in GaN/AlN nanowire and planar heterostructures, doped with Si and with Ge, with various doping levels. In the case of the GaN/AlN planar structures, ISB absorption around the 1.55 μm telecommunication wavelength is well documented in the literature [172]. The multi-Lorentzian-peak spectral profile observed in figures 20(a) and (b) is due to in-plane thickness fluctuations in the quantum wells [183,184]. This structure is spectrally resolved because the linewidth of the transitions is smaller than the energetic difference associated to increasing the quantum well thickness by one monolayer. The blueshift and broadening of the ISB transition with doping is assigned to many-body effects, specifically exchange interactions and plasmon screening or depolarization [185].

In the case of these nanowire heterostructures, the TM-polarized absorption corresponds to the transition between the ground electron level of the GaN well and the first excited electron level associated to confinement along the growth axis (s-$p_z$). The spectral shape is rather Gaussian, due to the inhomogeneous distribution of well thickness/diameter from wire to wire and along the nanowires. This prevents resolving monolayer thickness fluctuations in the axial direction, similar to observations in GaN/AlN quantum dots grown by the Stranski-Krastanov method [186]. Figure 20 presents the best result in the literature in terms of linewidth of ISB absorption in GaN nanowires, with a full width at half maximum of 200 meV. When increasing the doping level, the absorption blueshifts due to many-body effects. However, the linewidths do not change significantly since it is not dominated by the interaction with ionized impurities but by structural inhomogeneities within the nanowire ensemble.

Resonant tunneling transport has also been demonstrated in single nanowires containing GaN/AlN heterostructures [187,188]. In the case of a GaN/AlN double barrier in a single *n-i-n* GaN nanowire dispersed and contacted on $SiO_2$-on-silicon [187], features associated to NDR appeared at both negative and positive bias, and their location could be tuned by adjusting the electrostatic potential via a gate contact deposited on the back side of the silicon carrier wafer. In the case of multiple GaN/AlN wells [188], reproducible NDR associated to the electron tunneling through the quantum-confined electron states was also observed.

Using GaN/AlN (2 nm / 3 nm) nanowire heterostructures with ISB absorption around 1.55 μm, a single nanowire QWIP has been demonstrated [189]. The spectral response of the detector was obtained by measuring the photocurrent induced by various laser diodes operating at different wavelengths along the near-IR spectrum, as shown in figure 21. The response is maximum around 1.3-1.55 μm, which is consistent with ISB absorption from the first electron level in the wells to the first excited level associated to the confinement along the nanowire growth axis ($e_1$ to $e_{2z}$). Then, the response goes down at 1.0 μm and increases again at shorter wavelengths, which can be attributed to electron transitions to higher energy levels ($e_{3z}$, $e_{4z}$), theoretically predicted at 0.84 and 0.72 μm. Unlike the interband photocurrent (response to UV illumination in this case), the ISB photocurrent scales linearly with the incident optical power. This confirms that the ISB transitions are less sensitive to surface-related phenomena, as theoretically predicted [179]. The responsivity at 1.55 μm was estimated by assuming that the active area of the detector $A_{opt}$ corresponded to the surface nanowire that was exposed to the laser, obtaining values of 0.6±0.1 A/W and 1.1±0.1 A/W when measuring at chopping frequencies of 647 and 162 Hz, respectively.

# 5. Solar cells

One of the main challenges of our time is the evolution from energy systems based on fossil fuels towards a larger use of renewable resources. The deployment of solar cells is playing a crucial role in this energy transition. Huge research efforts are oriented to increase the solar cell power conversion efficiency, to increase the achievable "Watts per square meter" and therefore decrease the total system cost. Commercial solar cells are based on planar single junctions, whose efficiency is limited to about 33% according to the Shockley-Queisser theory [190,191]. In this context, nanowire junctions are attracting a lot of interest for the fabrication of solar cells [192–194] due to their increased lattice mismatch tolerance, which enables the implementation of tandem (multi-junction) solar cell designs with more flexibility than planar layers. Furthermore, the light concentration capabilities of nanowires make it possible to reduce the amount of active material without degradation of the absorption. Finally, nanowires are compatible with multiple substrates, such as silicon or flexible materials, which is interesting not only for large scale solar cells but also for the development of wearable devices. Due to the societal relevance of this research, we have decided to include this section to briefly review progress in this domain.

Various architectures have been considered for the fabrication of nanowire solar cells, namely radial and axial p-n junctions and radial type-II heterojunctions. So far, the efficiency of type-II heterojunctions is lower than that of p-n junction devices, and radial p-n junctions have shown relatively low open-circuit voltage, which points to a lower junction quality in comparison to axial junctions. As outstanding results, 11.1% [195] and 13.8% [196] conversion efficiencies were reported in 2013 by Eindhoven University of Technology and Lund University, respectively, both using axial-homojunction InP nanowire ensembles synthesized by localized growth in MOVPE using Au as a catalyst. In the same year, SunFlake A/S reported 10.2% conversion efficiency in a single-nanowire device consisting of a GaAsP radial homojunction (self-catalytic growth by MBE). In 2016, 15.3% conversion efficiency was reported by Sol Voltaics AB (spinoff company of Lund University) using an ensemble of axial homojunction GaAs nanowires (localized growth by MOVPE using Au as a catalyst) with an AlGaAs shell for surface passivation [197]. Finally, 17.8% efficiency was reported by Eindhoven University of Technology (also in 2016) in a homojunction InP nanowire solar cell fabricated through the top-down method [198].

To optimize the performance of the solar cells, efforts on doping, surface passivation and fabrication technology are still required, and the introduction of this technology in the market will necessitate also a reduction of the fabrication costs. The implantation of nanowire tandem architectures, still in a very exploratory stage [199], might also lead to a breakthrough in terms of power conversion efficiency.

# 6. Conclusions and perspectives

This paper presents an overview of various architectures under consideration for the fabrication of semiconductor nanowire photodetectors. We have discussed the advantages of nanowire photodetectors over planar technologies, including the extreme miniaturization, reduction of the absorbing material without detrimental effects on the absorption efficiency, huge responsivity, compatibility with silicon substrates and possibility of integration with flexible circuits. Nanowire photodetectors present extremely promising features, but some technological bottlenecks require further effort in order to improve performance, reproducibility, reliability and fabrication yield, while moderating the production costs. Main issues remaining are for example problems related to surface states, traps and defects, controlled doping, homogeneity of ternary and quaternary alloys, chemical sharpness of the junctions and heterostructures, high series and contact resistance and noise level. Uniformity between different nanowire photodetectors needs to be improved for industrial applications. Furthermore, being able to dope both p- and n- type in the same material system, without polytypisms, would significantly enhance the application prospects. These issues are to be solved in the coming years with further experimental and theoretical insight into the nanowire growth process, and the role of surfaces and passivation. New planarization and contact schemes should be developed to reach the stability and reproducibility requirements of a market product. Rigorous aging tests, extreme temperatures, shock tests should be done to provide insights on the feasibility of implementing nanowire photodetectors industrially. Compatibility with the current CMOS- and Si-based industry should be focused on, in view of the potential integration of the photodetector with the readout electronics. Finally, strategies to reduce costs are essential to place nanowire photodetectors in a competitive position with well-established technologies. In spite of these challenges, nanowire photodetectors are called to become keystone elements in the emerging fields of high-resolution imaging, quantum optics, integrated photonics, and high-speed optical signal processing, in addition to their potential as photochemical transducers.

**Acknowledgements:** We acknowledge funding from AGIR 2016 proposed by the community of Université Grenoble Alpes and the T-ERC ANR project e-See.

the band structure of group-III nitrides *Phys. Rev. B* **90** 125118

# Tables

Table 1. Some material parameters of wurtzite semiconductors.

| Parameters (units) | Symbol | GaN | AlN | InN | ZnO | MgO(*) | CdO(*) |
|---|---|---|---|---|---|---|---|
| Lattice constants (nm) | $a$ | 0.3189 [200] | 0.3112 [200] | 0.3545 [200] | 0.3166 [201] | 0.3221 [201] | 0.3678 [202] |
| | $c$ | 0.5185 [200] | 0.4982 [200] | 0.5703 [200] | 0.5070 [201] | 0.504 [201] | 0.5825 [202] |
| Spontaneous polarization (C.m$^{-2}$) | $P_{SP}$ | -0.034 [203] | -0.090 [203] | -0.042 [203] | -0.053 [201] | -0.080 [201] | -0.047 [202] |
| Piezoelectric constants (C.m$^{-2}$) | $e_{31}$ | -0.49 [204] | -0.60 [204] | -0.57 [204] | -0.55 [201] | -0.78 [201] | -0.48 [202] |
| | $e_{33}$ | 0.73 [204] | 1.46 [204] | 0.97 [204] | 1.24 [201] | 0.14 [201] | 1.67 [202] |
| | $e_{15}$ | | | | -0.38 [201] | -0.36 [201] | |
| Elastic constants (GPa) | $c_{11}$ | 390 [205] | 396 [206] | 223 [206] | 238 [201] | 205 [201] | 150 [202] |
| | $c_{12}$ | 145 [205] | 137 [206] | 115 [206] | 106 [201] | 80 [201] | 108 [202] |
| | $c_{13}$ | 106 [205] | 108 [206] | 92 [206] | 84 [201] | 88 [201] | 105 [202] |
| | $c_{33}$ | 398 [205] | 373 [206] | 224 [206] | 176 [201] | 222 [201] | 61 [202] |
| | $c_{44}$ | 105 [205] | 116 [206] | 48 [206] | 58 [201] | 58 [201] | 47 [202] |
| Band gap (eV) | $E_G$ | 3.51 [200] | 6.2 [200] | 0.69 [47] | 3.43 [207] | 6.4 [207] | 0.92 [207] |
| Electron effective mass | $m_e^*$ | 0.2 [200] | 0.3 [208] | 0.07 [200] | 0.24 [209] | 0.37 [209] | 0.16 [209] |
| Deformation potentials (eV) $a_{cz}$, $a_{ct}$ = conduction band deformation potentials $D_i$ = valence band deformation potentials | $a_{cz}$-$D_1$ | -5.81 [210] | -4.3 [210] | -3.62 [210] | -3.06 [209] | -1.95 [209] | -2.81 [209] |
| | $a_{ct}$-$D_2$ | -8.92 [210] | -12.1 [210] | -4.60 [210] | -2.46 [209] | -7.96 [209] | -0.29 [209] |
| | $D_3$ | 5.47 [210] | 9.12 [210] | 2.68 [210] | 0.47 [209] | 5.87 [209] | -1.86 [209] |
| | $D_4$ | -2.98 [210] | -3.79 [210] | -1.74 [210] | -0.84 [209] | -1.97 [209] | -0.30 [209] |
| | $D_5$ | -2.82 [210] | -3.23 [210] | -2.07 [210] | -1.21 [209] | -1.93 [209] | -0.91 [209] |
| | $D_6$ | - | - | - | -1.77 [209] | -3.03 [209] | -1.21 [209] |

**Figure captions**

**Figure 1.** (a) Schematic of optical concentration in a nanostructure, illustrating the difference between optical area and geometrical area. [Reprinted with permission from [6] Xu et al. 2015 *Scientific Reports* **5** 13536. © 2015 Springer Nature Ltd.]. (b) Absorption enhancement in a single vertical GaAs nanowire standing on a silicon substrate, as a function of the nanowire diameter ($2r$). [Reprinted with permission from [19] Heiss, et al. 2014 *Nanotechnology* **25** 014015. © 2014 IOP Publishing Ltd.]

**Figure 2.** (a) Schematic description of the VLS growth method. (b,c) Scanning electron microscopy images of GaAs nanowires grown by the VLS method using two different V/III beam-equivalent-pressure ratios. (d) Bright-field TEM image (scale bar = 10 nm) of an individual GaAs nanowire from (c). The wire has wurtzite crystal structure, as shown on the electron diffraction pattern (e). The region directly below the gold particle features a "cooling neck" with zinc-blende phase. [Reprinted with permission from [30] Beznasyuk, et al. 2017 *Nanotechnology* **28** 365602. © 2017 IOP Publishing Ltd.]

**Figure 3.** High-resolution transmission electron microscopy images illustrating the initial steps of the growth of GaN nanowires on Si(111) using an AlN buffer layer by PAMBE. Images were taken after 6, 9, 10 and 15 min of GaN growth under nitrogen-rich conditions: (a) spherical-cap-shaped island with an inset representing a high magnification of the first AlN monolayers at the interface, (b) truncated-pyramid-shaped island, (c) full-pyramid-shaped island, and (d) nanowire. The pyramid-shaped islands and the nanowire are hexahedral. All the shapes are outlined for the sake of clarity. [Reprinted with permission from [35] Consonni, et al. 2010 *Phys. Rev. B* **81** 085310. © 2010 The American Physical Society]

**Figure 4.** GaN nanowires generated by selective area growth using a Ti mask. (a-c) Scanning electron microscopy images of regular triangular-lattice arrays of GaN nanowires with a 500 nm period. The wire diameters are (a) 122 nm, and (b,c) 170 nm. (d-f) Bright-field transmission electron microscopy images of the GaN nanowires synthesized by selective area growth on a GaN-on-Si(111) structure. (d) View of the epitaxial layers including the Si substrate, an AlN/GaN superlattice (SL) buffer layer, a GaN thick layer and the GaN rods. (e) Nanowire with a diameter of 122 nm. (f) Magnified image of the bottom of the nanowire in (e). [Reprinted with permission from [42] Kishino et al. 2015 *Nanotechnology* **26** 225602. © 2015 IOP Publishing Ltd.]

**Figure 5.** (a) Top-down fabrication process of GaN nanowires using silica colloids as semi-periodic mask (b-e) Cross sectional SEM images showing dry-etched GaN posts transiting into GaN nanowires (b) before wet etch, (c) after 2 hours, (d) after 6 hours and (e) after 9 hours from start of wet etch. All images have the same magnification. Scale bars are 2 μm. [Reprinted with permission from [45] Li et al. 2012 *Optics Express* **20** 17874. © 2012 IOP Publishing Ltd.]

**Figure 6.** (a) Schematic description of the wurtzite lattice, outlining the (0001) *c*-plane, (11-20) *a*-plane and (-1-100) *m*-plane. (b) Band diagram of a GaN QW in an infinity GaN/AlN (2.1 nm / 3 nm) MQW system. In gray lines, squared wavefunctions associated to the first and second confined levels of electrons ($e_1$, $e_2$) and the first confined level of holes ($h_1$). (c) Type I heterostructure Band offsets between binary wurtzite compounds in the AlGaInN and MgZnCdO systems. The band gap energies are given at low temperature. Note that the most stable configuration of MgO and CdO is cubic. Therefore, the values presented here are the result of first-principle calculations. For nitrides, the conduction band offsets are ≈ 1.8 eV for AlN/GaN [50] and ≈ 2.2 eV for GaN/InN [51], and for oxides, they are calculated from the valence band offsets in ref. [52]: ≈ 1.2 eV for MgO/ZnO and ≈ 0.1 eV for ZnO/CdO (d) Band offsets in type-II ZnO/ZnX (X = S, Se, Te) heterostructures. Offset values are taken from ref. [53]. The bandgap energies are given at room temperature, with each material in its more stable crystallographic configuration (wurtzite for ZnO and zinc-blende for ZnS, ZnSe and ZnTe).

**Figure 7.** Transmission electron microscopy images of (a) spontaneously grown and (d) catalyst-assisted VLS grown ZnO nanowires, together with (b,e) experimental and (c,f) simulated convergent electron beam diffraction patterns, which indicate (a-c) Zn-polar and (d-f) O-polar growth. [Reprinted with permission from [66] Sallet, et al. 2013 *Appl. Phys. Lett.* **102** 182103. © 2013 AIP Publishing]

**Figure 8.** Schematic description of (a) a single-nanowire photodetector and (b) a nanowire ensemble photodetector.

**Figure 9.** (a) Scanning electron microscopy image of n-i-n GaN nanowires on a Si(111) substrate. Insets: Top panel is the top-view of a contacted single nanowire; bottom panel shows the measurement scheme for characterization of a single nanowire device. (b) Photocurrent variation from a single GaN nanowire as a function of the excitation power (λ = 244 nm) measured with a synchronous detection system at various light chopping frequencies. Dashed lines are fits to the relation $I_{ph} \propto P_{opt}^{\beta}$ using the values of $\beta$ indicated in the legend. (c) Spectral response of a single GaN nanowire photodetector under 3 V bias measured at various light chopping frequencies. The spectra were corrected by the lamp intensity taking the measurements of the degree of linearity in (b) into account. [Reprinted with permission from [70] González-Posada, et al. 2012 *Nano Lett.* **12** 172–6. © 2012 American Chemical Society]

**Figure 10.** (a) Current-voltage characteristics of a ZnO single nanowire photodetector as a function of the light intensity. From top to bottom, the light intensity was $4\times10^{-2}$ W/cm$^2$ (black), $4\times10^{-3}$ W/cm$^2$ (red), $4\times10^{-4}$ W/cm$^2$ (green), $1.3\times10^{-4}$ W/cm$^2$ (blue), $4\times10^{-5}$ W/cm2 (cyan), $1.3\times10^{-5}$ W/cm2 (magenta), $6.3\times10^{-6}$ W/cm$^2$ (yellow), and in dark (brown). Inset is the SEM image of a typical ZnO nanowire device. After growth, the nanowires were dispersed onto a thermally oxidized Si substrate (600 nm SiO$_2$), and consequently, Ti/Au (20 nm/160 nm) interdigitated electrodes with 2 µm finger spacing were patterned on top of the nanowires using optical lithography. [Reprinted with permission from [73] Soci, et al. 2007 *Nano Lett.* **7** 1003-9. © 2007 American Chemical Society]

**Figure 11.** Schematic description of the conduction ($E_C$) and valence ($E_V$) band profile along the nanowire diameter in the dark (black) and under illumination (grey) for nanowires that are (a) partially depleted, (b) at the threshold of full depletion in the dark, and (c) fully depleted. The dash-dotted line marks the Fermi level in the dark ($E_F$). In all the cases, the Fermi level is considered pinned at the nanowire sidewalls at an energy position $E_C - \Delta E$. Shadowed areas mark the depletion region in the dark and under illumination.

**Figure 12.** (a) Variation of $\beta$ as a function of the dark current (measured at +1 V) in single GaN nanowires. Note, the correlation of almost-linear nanowires with dark current in the nA range (group A), and clearly sublinear nanowires with dark current in the µA range (group B). The solid line is a guide to the eye. Inset: dark current at +1 V bias as a function of the diameter of the nanowires measured by scanning transmission electron microscopy. The dotted line is a guide to the eye. (b) Spectral response measurements for typical group-A and group-B nanowire specimens. The dashed line marks the wavelength

of the GaN band gap at room temperature ($E_{g\,GaN}$). [Reprinted with permission from [93] Spies, et al. 2018 *Nanotechnol.* **29** 255204. © 2018 IOP Publishing Ltd.]

**Figure 13.** Schematic description of various p-n junction configurations involving nanowires: (a) radial (core-shell), (b) axial, (c) crossing n and p nanowires, (d) branched p-n junction, and n-type nanowires on p-type substrate.

**Figure 14.** Schematic of an InP/InAsP nanowire avalanche photodetector, along with energy-dispersive x-ray spectroscopy linescans superimposed on a transmission electron microscopy image. (b) Current-voltage characteristics and gain at 77 K of an InP/InAsP nanowire avalanche photodetector in the dark, before and after annealing, and after selective illumination of the InAsP absorption region. [Reprinted, with permission, from [139] Jain, et al. 2017 *ACS Photonics* **4** 2693. © 2017 American Chemical Society]

**Figure 15.** Schematic description of (a) radial (core-shell) and (b) axial nanowire heterostructures.

**Figure 16.** (a) Schematic description of the structure consisting of a GaN nanowire with an AlN axial insertion. The lateral growth during the deposition of the AlN and top GaN sections results in the formation of a GaN/AlN/GaN core/shell structure at the base of the nanowire. The current (red arrows) drifts towards the nanowire surface. (b) Scanning transmission electron microscopy image of a contacted nanowire and (c) magnified image of the AlN insertion. The AlN shell is visible in the image as a dark contrast. (d) Variation of the dark current and photocurrent as a function of the measurement atmosphere. The photocurrent is induced with illumination of ≈ 0.2 W/cm² of UV light (λ = 244 nm). [Reprinted with permission from [165] den

Hertog, et al. 2013 *Jpn. J. Appl. Phys.* **52** 11NG01. © 2013 The Japan Society of Applied Physics]

**Figure 17.** (a) Sketch of the nanowire structure under study. (b), (c) High angle annular dark field (HAADF) scanning transmission electron microscopy (STEM) images of two coalesced nanowires with different magnification. The bright and dark gray areas correspond to GaN and AlN, respectively. (d) Photocurrent ($I_{ill}$) and photosensitivity factor ($I_{ill}/I_{dark}$) for different nanowires plotted as a function of $I_{dark}$. Both $I_{ill}$ and $I_{dark}$ are measured at 1 V bias. The value of $I_{ill}$ corresponds to an excitation density of ≈ 1 W/cm$^2$ ($\lambda$ = 325 nm). [Reprinted with permission from [89] Lähnemann, et al. 2016 *Nano Lett.* **16** 3260. © 2016 American Chemical Society]

**Figure 18.** (a) Schematic description of a GaN nanowire containing a GaN/AlN multi-quantum-well heterostructure. The voltage convention is indicated with bias applied to the GaN cap ($V_B$) whereas the GaN stem is grounded (GND). (b) Spectral response of a single nanowire at various values of bias voltages (±0.5, ±1, and ±2 V). The nanowire structure and measuring circuit are as described in (a). Data are corrected by the Xe-lamp emission spectrum taking the sublinear power dependence of the nanowire into account. Measurements are normalized and vertically shifted for clarity. Solid (dashed) lines correspond to positive (negative) bias. Shadowed areas outline the difference in the response between positive and negative bias. (c-e) Description of the evolution of the band diagram under bias: (c) negative bias, (d) zero bias, and (e) positive bias. The data is the result of one-dimensional calculations of the nanowire heterostructure in the presence of an external electric field. As the model is one-dimensional, surface states are not taken into account in these calculations. The Fermi level(s) is (are) indicated by a dash-dotted blue line in each diagram. [Reprinted with permission from [167] Spies, et al. 2017 *Nano Lett.* **17** 4231. © 2017 American Chemical Society]

**Figure 19.** Top: High angle annular dark-field (HAADF) scanning transmission electron microscopy (STEM) images of a Ge-doped ([Ge] = 3×10$^{19}$ cm$^{-3}$) GaN/AlN nanowire at two different magnifications. The heterostructure consists of 30 periods of Ge-doped GaN/AlN (2 nm/3 nm) quantum wells. Below the heterostructure, the GaN nanowire base is 800 nm long with a diameter of 60 nm, and the heterostructure is capped with 30 nm of GaN. Dark (bright) contrast corresponds to AlN (GaN). Bottom: High-resolution x-ray diffraction θ–2θ scan of the (0002) reflection of planar and nanowire GaN/AlN heterostructures, doped with Si or Ge. The diffractograms are normalized to the maximum of the GaN reflection and vertically shifted for clarity. The peak labeled SL corresponds to the main reflection of the GaN/AlN superlattice. [Reprinted, with permission, from [181] Ajay, et al. 2017 *physica status solidi (b)* **254** 1600734. © 2017 WILEY-VCH Verlag GmbH & Co.]

**Figure 20.** Room-temperature absorption spectra for TM-polarized light measured in (a) Si-doped planar heterostructures (PS1-PS3), (b) Si-doped nanowire heterostructures (NS1, NS2), (c) Ge-doped planar heterostructures (PG1-PG3), and (d) Ge-doped nanowire heterostructures (NG1, NG2). The nominal thickness of the heterostructures is 2 nm GaN / 3 nm AlN, repeated 30 times. Samples PS1-PS3 and NS1-NS2 are doped with Si in the GaN wells, with concentrations PS1, NS1: [Si] = 3×10$^{19}$ cm$^{-3}$; PS2, NS2: [Si] = 1×10$^{20}$ cm$^{-3}$; PS3: [Si] = 3×10$^{20}$ cm$^{-3}$. Samples PG1-PG3 and NG1-NG2 are doped with Ge in the GaN wells, with concentrations PG1, NG1: [Ge] = 3×10$^{19}$ cm$^{-3}$; PG2, NG2: [Ge] = 1×10$^{20}$ cm$^{-3}$; PG3: [Ge] = 3×10$^{20}$ cm$^{-3}$. The spectra are normalized to their maximum and vertically shifted for clarity. Theoretical values combining 1D nextnano[3]

and many-body calculations are indicated as arrows in (a), (c). Theoretical values from 3D nextnano3 calculations are indicated as arrows in (b), (d). [Reprinted, with permission, from [180] Ajay, et al. 2017 *Nanotechnology* **28** 405204. © 2017 IOP Publishing Ltd.]

**Figure 21.** Schematic representation of the single-nanowire QWIP containing a GaN/AlN heterostructure. Conduction band (CB) diagram of one of the GaN wells in the heterostructure, indicating the grown electron level ($e_1$) and the excited levels associated to electron confinement along the growth axis ($e_{2z}$, $e_{3z}$, $e_{4z}$). The levels are represented with their respective squared wavefunctions. The horizontal dashed line represents the fermi level ($E_F$). On the right side, photocurrent as a function of the excitation wavelength measured in two different devices (NW 1 and NW 2). The spectra are normalized to the response at 1.3 μm. [Reprinted, with permission, from [189] Lähnemann, et al. 2017 *Nano Letters* **17** 6954. © 2017 American Chemical Society]

**Figure 1**

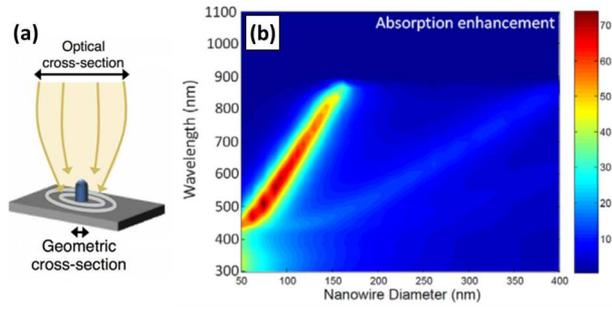

**Figure 2**

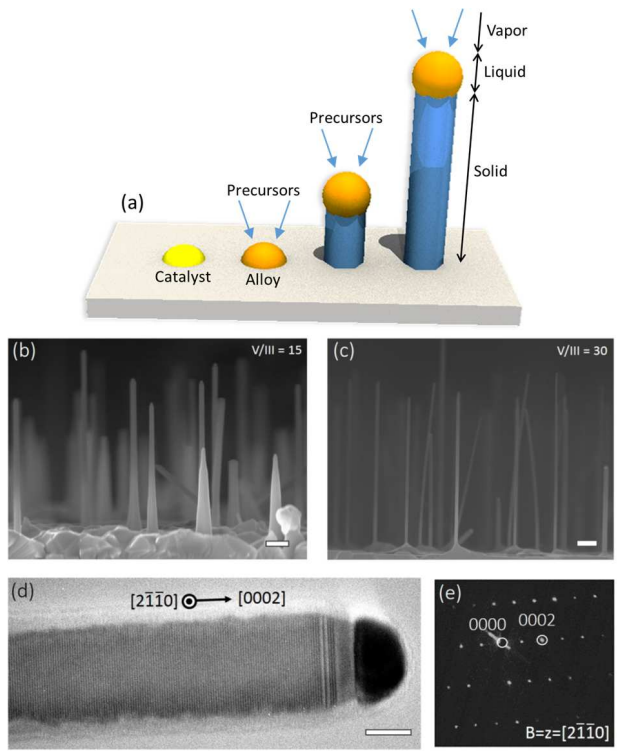

**Figure 3**

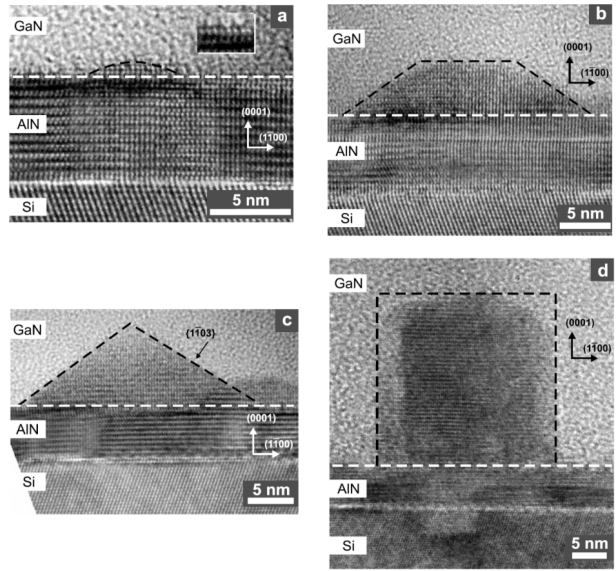

**Figure 4**

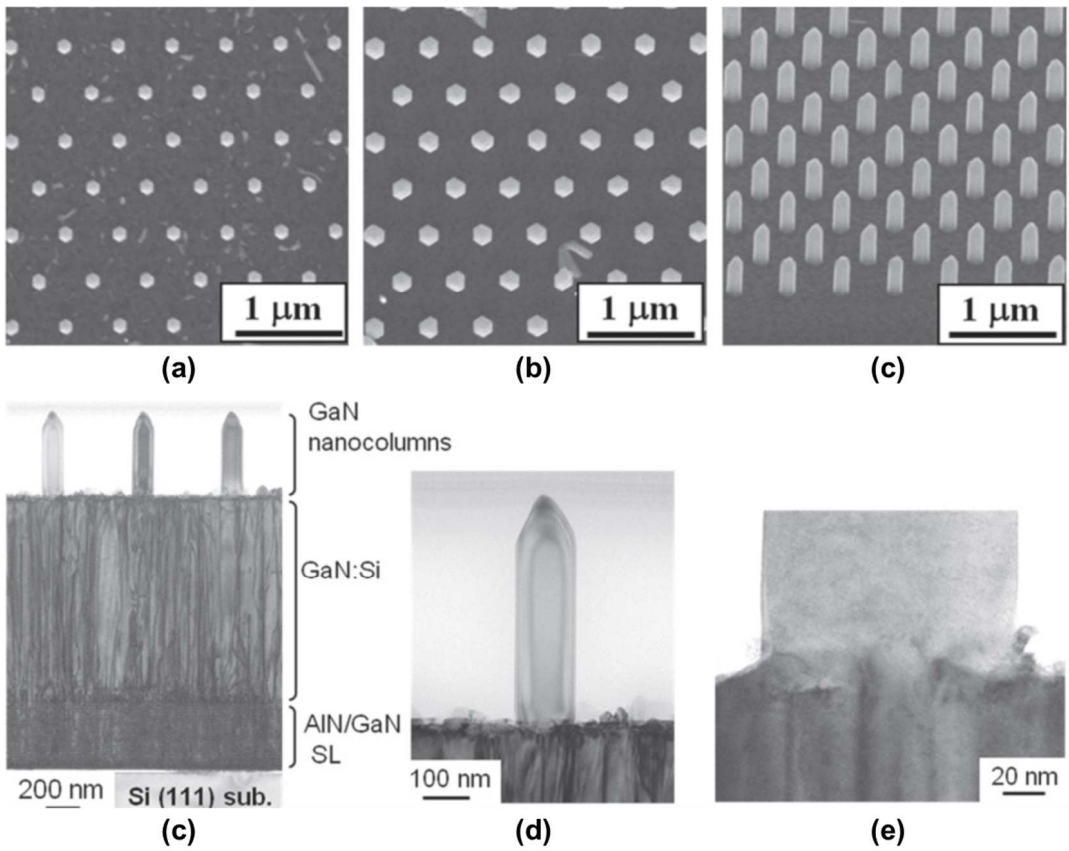

**Figure 5**

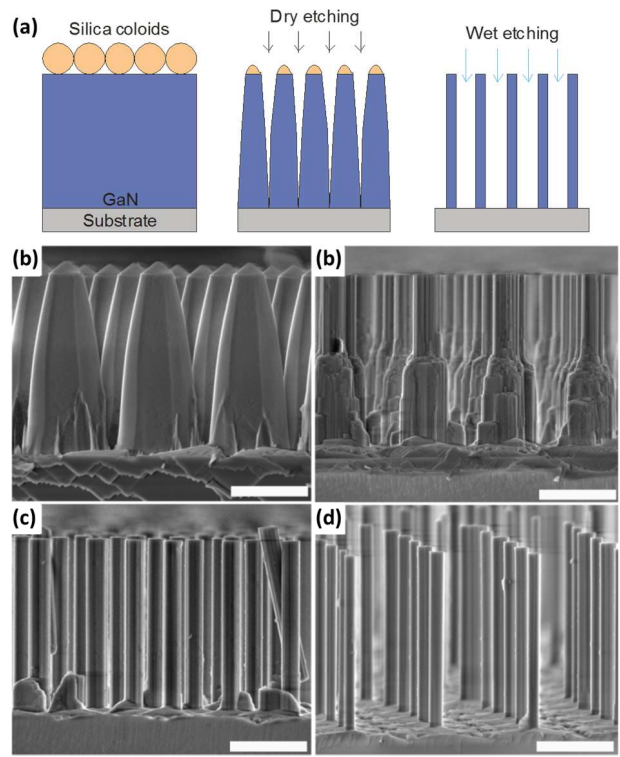

**Figure 6**

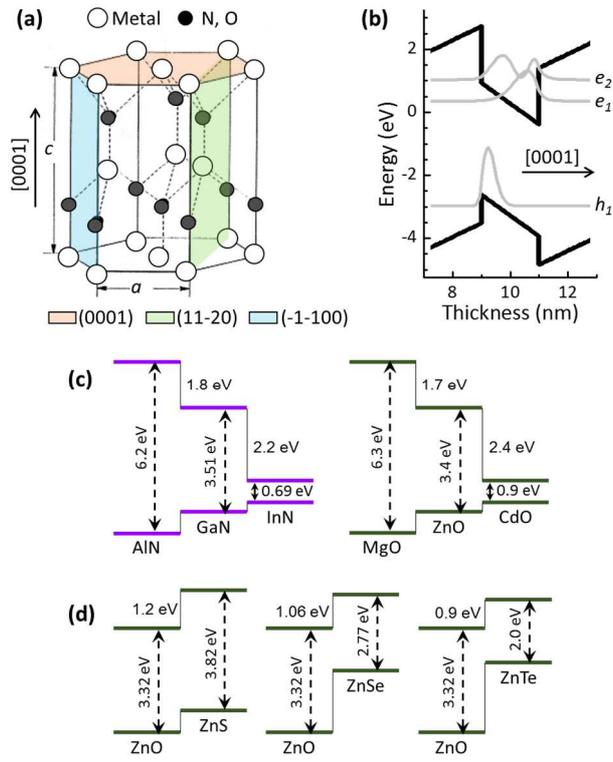

**Figure 7**

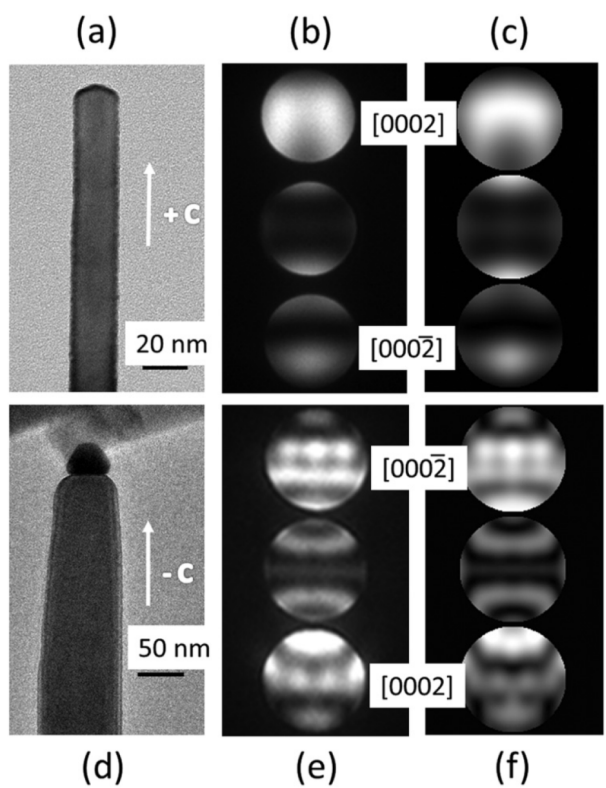

**Figure 8**

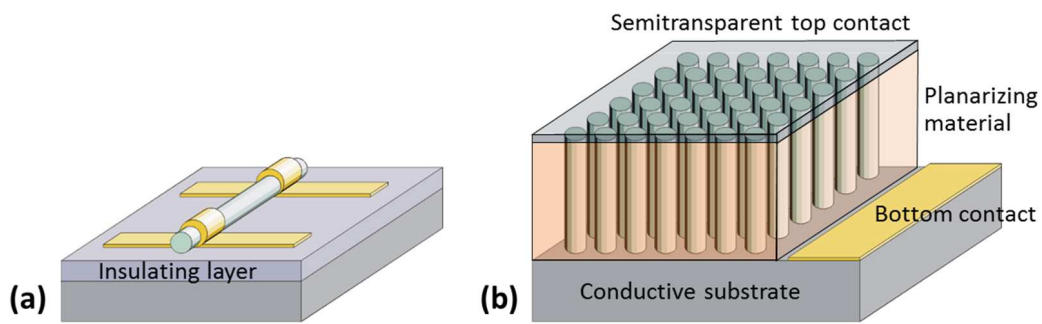

**Figure 9**

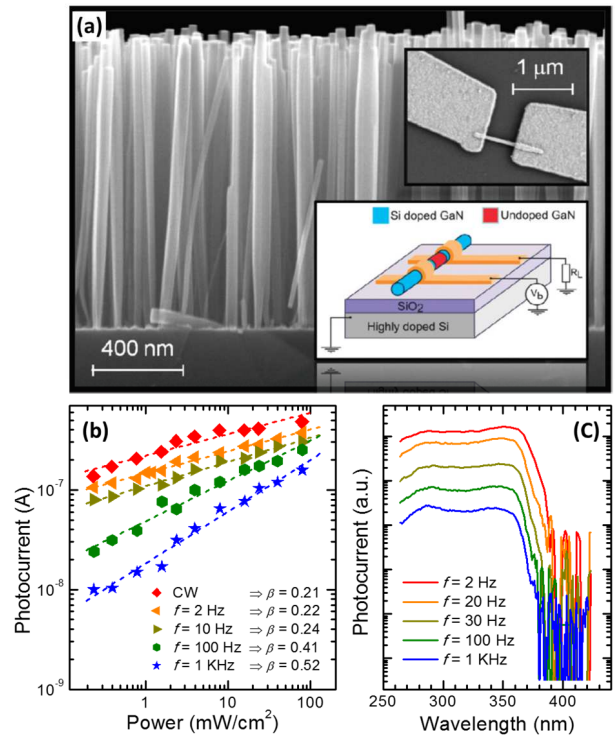

**Figure 10**

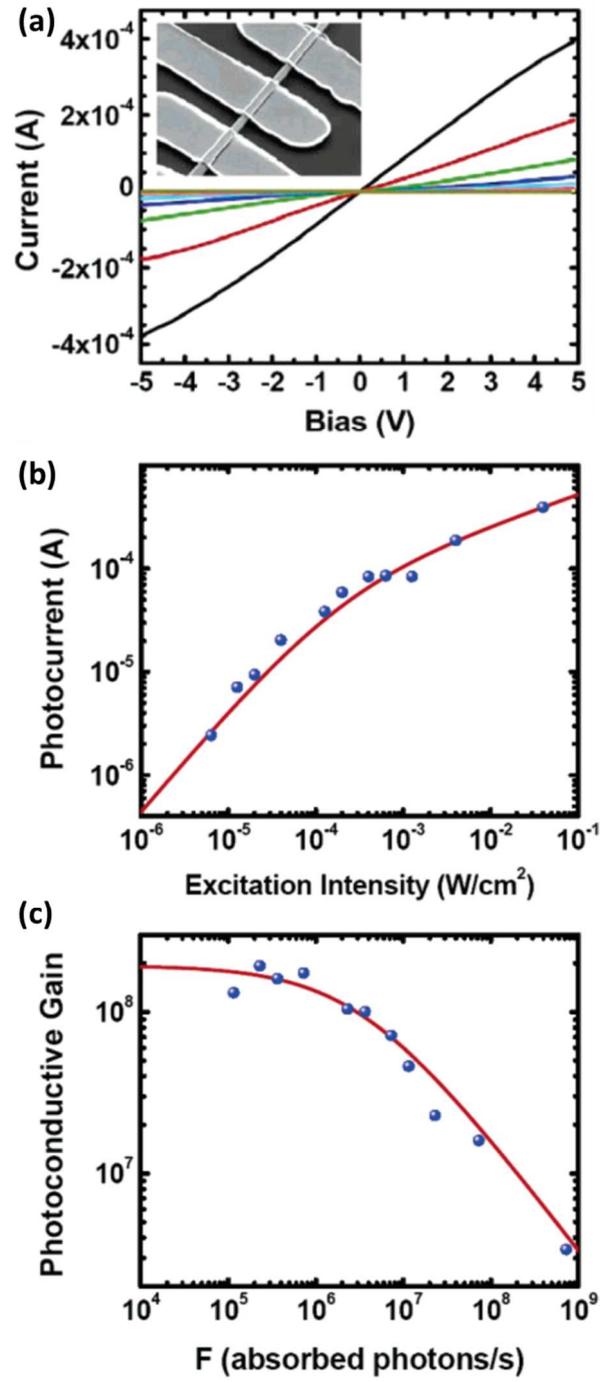

# Figure 11

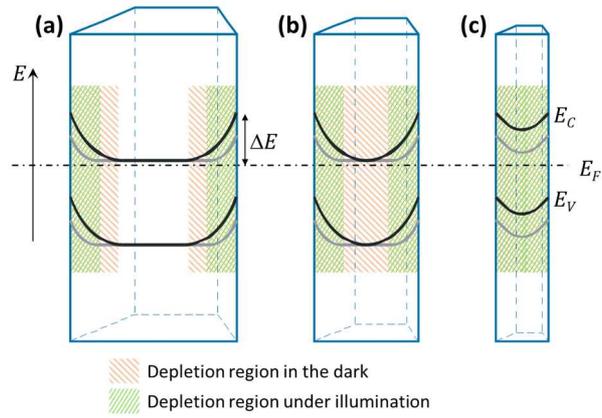

# Figure 12

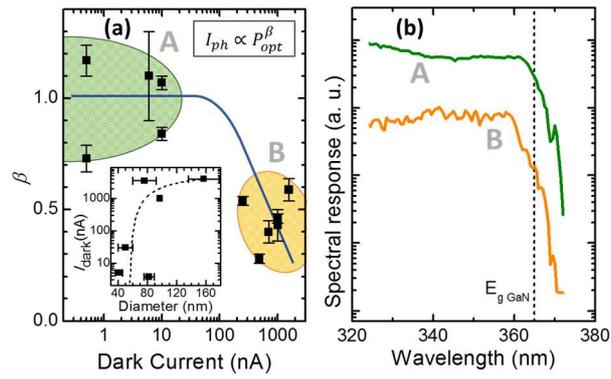

**Figure 13**

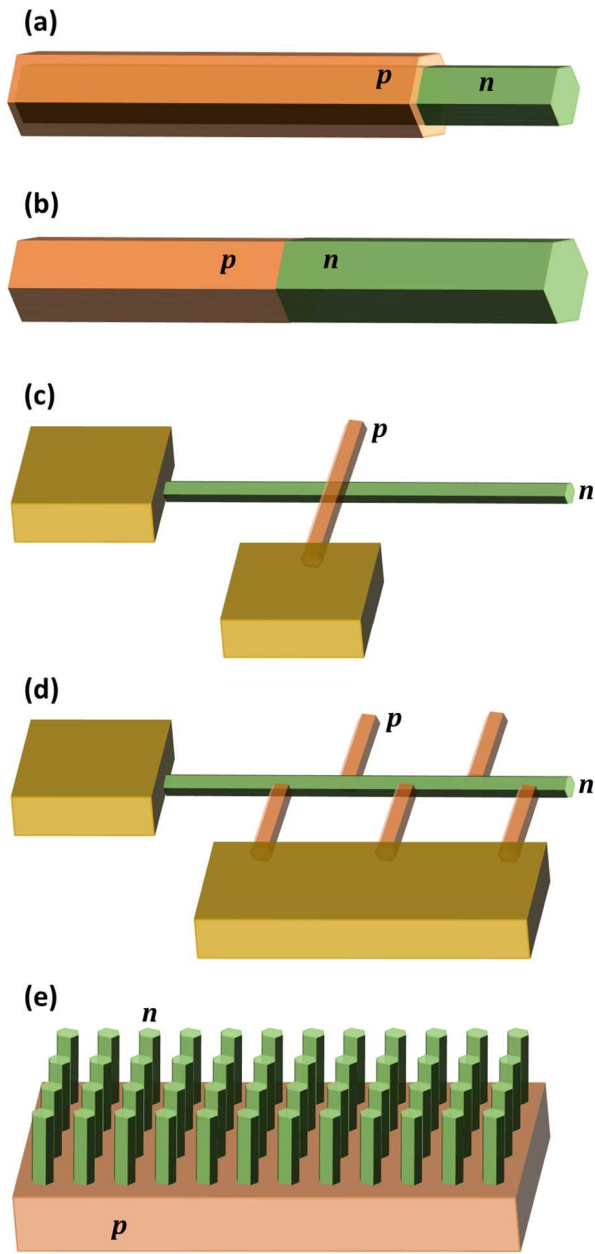

**Figure 14**

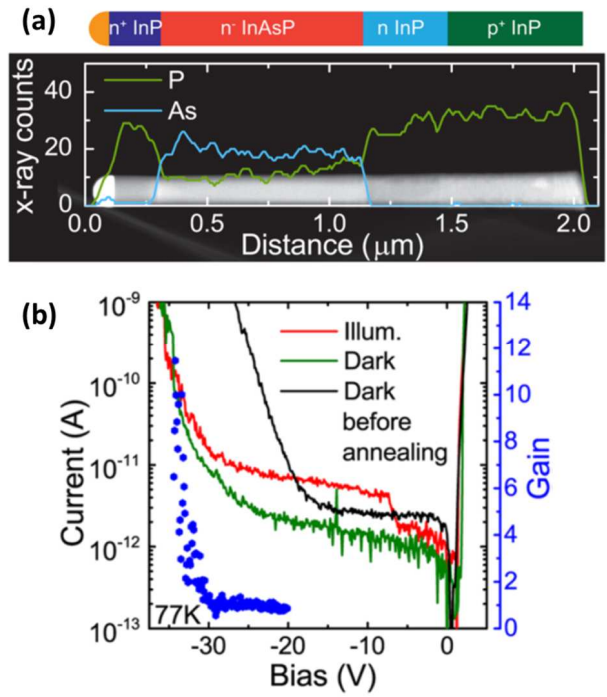

**Figure 15**

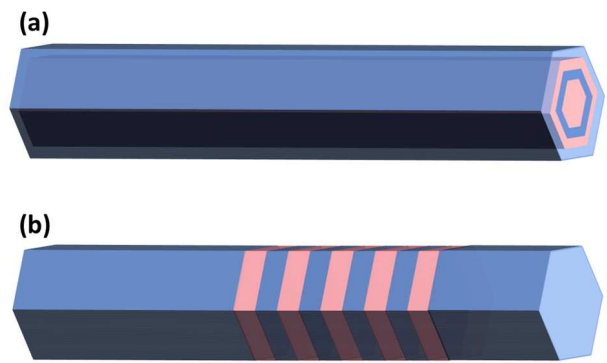

**Figure 16**

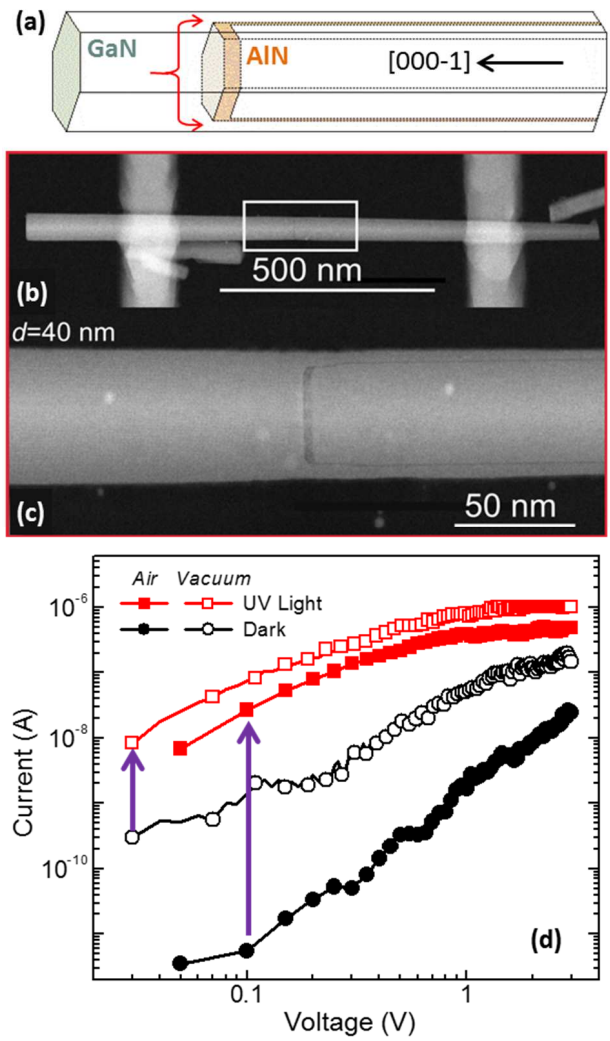
75

**Figure 17**

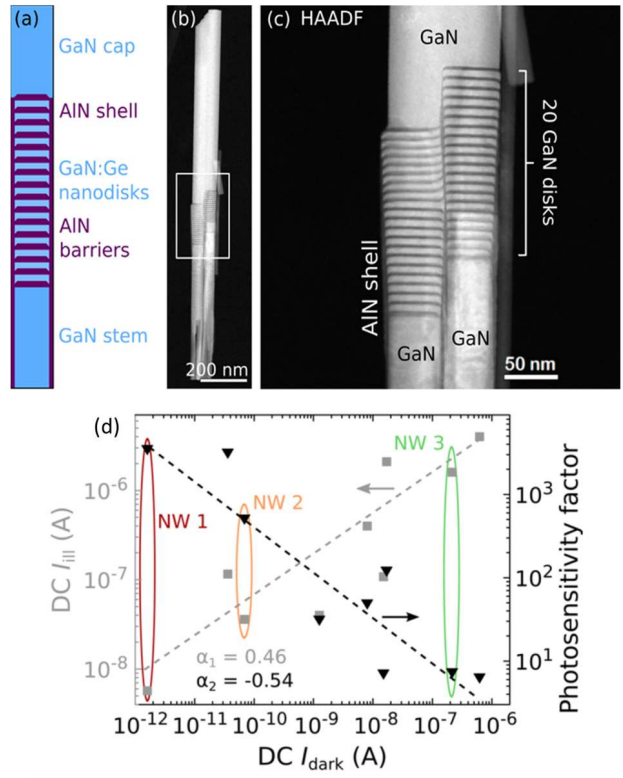

**Figure 18**

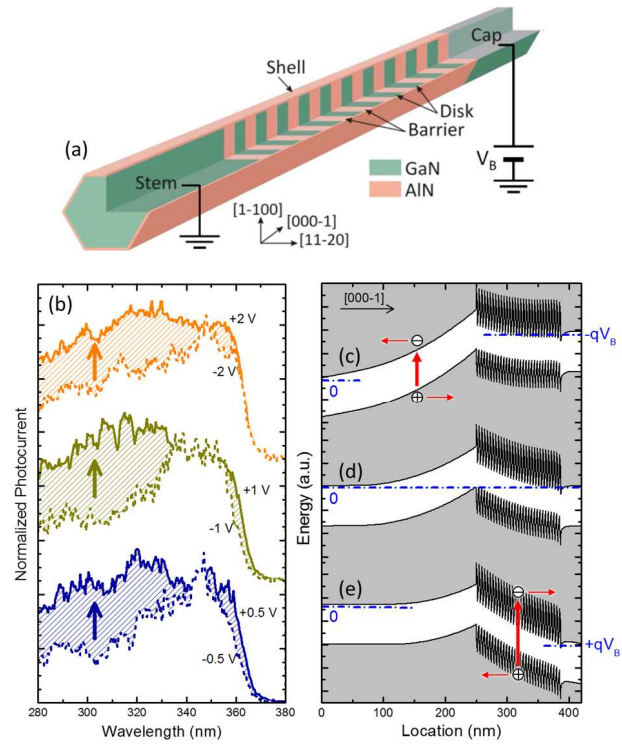

**Figure 19**

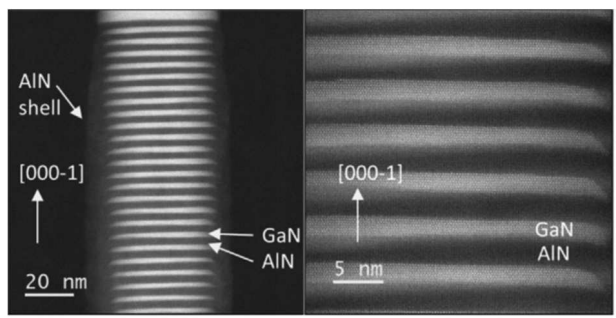

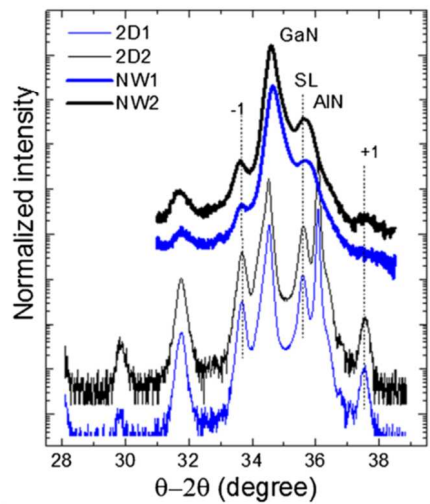

**Figure 20**

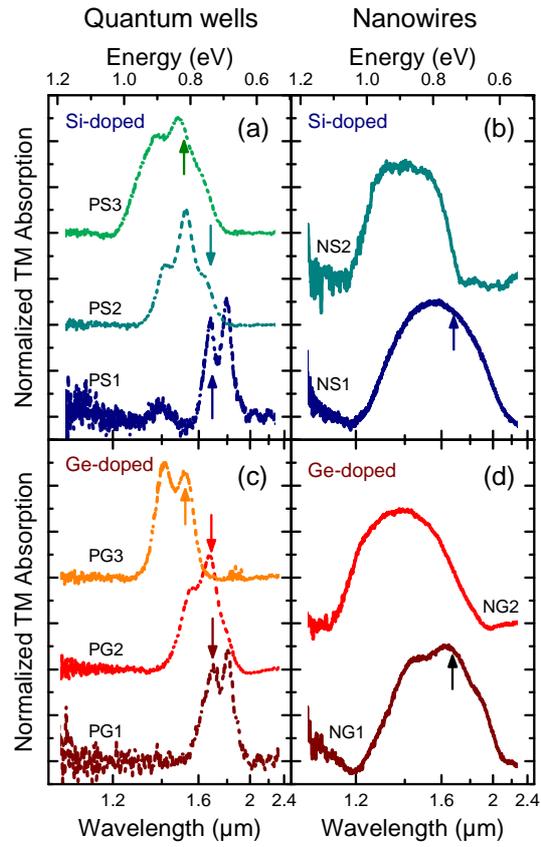

**Figure 21**

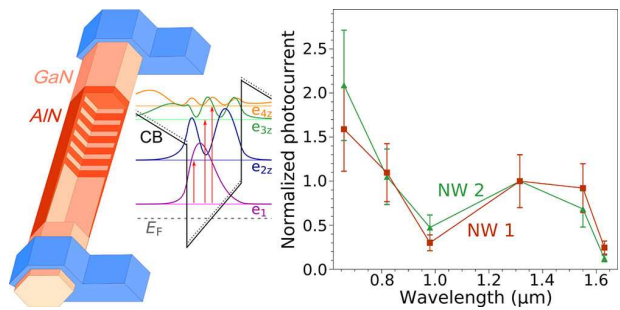